\documentclass[a4paper,12pt]{article}
\usepackage{jcappub}
\usepackage{graphicx}
\usepackage{dcolumn}
\usepackage{bm}
\usepackage{physics}
\usepackage{hyperref}
\hypersetup{
     colorlinks=true,
     linkcolor=blue,
     filecolor=blue,
     citecolor = blue,      
     urlcolor=blue,
     }
\usepackage[super]{nth}
\usepackage{physics}

\usepackage[shortlabels]{enumitem}
\usepackage{subcaption}
\usepackage{bbm}
\usepackage{orcidlink}
\numberwithin{equation}{section}
\usepackage{multicol}

\newcommand{\gi}[1]{\ensuremath{g^{(#1)}_{\mu\nu}}}
\newcommand{\Rsi}[1]{\ensuremath{R_{(#1)}}}
\newcommand{\Wi}[1]{\ensuremath{W^{(#1)\mu}_{\;\;\;\;\;\;\nu}}}
\newcommand{\Gi}[1]{\ensuremath{G^{(#1)\mu}_{\;\;\;\;\;\;\nu}}}

\newcommand{\Tmunui}[1]{\ensuremath{T^{(#1)\mu}_{\;\;\;\;\;\;\nu}}}
\newcommand{\vbein}[3]{\ensuremath{e_{#2}^{(#1)#3}}}
\newcommand{\ivbein}[3]{\ensuremath{e_{\;\;\;\;\;#3}^{(#1)#2}}}

\newcommand{\Mp}{\ensuremath{M_{\text{Pl}}}}

\newcommand{\Tcoeffs}{\ensuremath{T_{i_1\hdots i_D}}}
\newcommand{\Tz}{\ensuremath{T_{iiii}}}
\newcommand{\Ti}{\ensuremath{T_{iii,i+1}}}
\newcommand{\Tii}{\ensuremath{T_{ii,i+1,i+1}}}
\newcommand{\Tiii}{\ensuremath{T_{i,i+1,i+1,i+1}}}

\newcommand{\dy}{\ensuremath{\delta y}}

\newcommand{\hodge}{\ensuremath{\star}}
\newcommand{\GC}[1]{\ensuremath{\text{GC}_{#1}}}

\begin{document}

\title{A new look at multi-gravity and dimensional deconstruction}
\author[1]{Kieran Wood \orcidlink{0000-0002-4680-5563}}
\emailAdd{kieran.wood@nottingham.ac.uk}
\affiliation[1]{Nottingham Centre of Gravity, School of Physics and Astronomy,\\ University Of Nottingham, Nottingham, NG7 2RD, UK}

\abstract{
    It has long been understood that certain theories of ghost free massive gravity and their multi-graviton extensions can be thought of as arising from a higher dimensional theory of gravity, upon discretising the extra dimension. However, this correspondence between standard multi-gravity and extra dimensional gravity holds only when one discretises the extra dimension \emph{after} gauge fixing the lapse function associated to the various lower dimensional hypersurfaces. The lapse provides crucial structure to the extra dimensional theory: in pure general relativity (GR), it ensures full diffeomorphism invariance of the theory, and enforces its Hamiltonian constraint. Thus, upon deconstruction, important information related to the extra dimension is missing in the resulting multi-gravity theory; as a result one could never hope to recover higher dimensional GR in its entirety upon taking the appropriate continuum limit. Here, we develop an improved deconstruction procedure that maintains the free lapse, and show that the resulting deconstructed theory is essentially multi-gravity equipped with additional dynamical scalar fields, whose field equations encode the Hamiltonian constraint in the extra dimensional theory. As an example, we explicitly demonstrate that -- with an FLRW ansatz for the metrics in this new theory -- one may recover all of the equations and constraints of 5-dimensional brane cosmology upon taking the continuum limit. We then treat the deconstructed theory as an entity in its own right, and generalise it to arbitrary dimension and interaction structures beyond those admitting a well-defined continuum limit. We dub this theory `scalar-tensor multi-gravity', and show that the new scalar equations change the structure of some simple solutions that were previously allowed in standard multi-gravity, in a manner that exactly mirrors what we expect from higher dimensional GR.
}

\maketitle
\newpage

\section{Introduction}

It is an old and familiar story, dating back to the time of Kaluza and Klein (KK), that the gravitational physics of any higher dimensional spacetime, whenever one or more of the extra dimensions is compact, manifests itself to a lower dimensional observer through the appearance of towers of massive spin-2 states called massive gravitons. The number of such states in the effective lower dimensional description is dictated by the requirement that the higher dimensional theory be valid up to its corresponding Planck scale; from an effective field theory (EFT) perspective, the heaviest KK graviton existing within the spectrum of the lower dimensional theory must have mass ${m\lesssim \Mp^{(D)}}$. A great many models of beyond the Standard Model physics, introduced to provide solutions to numerous outstanding fundamental problems, base themselves on the existence of compact extra dimensions (string theory being the obvious example, but there are others too: for example, the Randall-Sundrum braneworld models \cite{RS1,RS2} that we will talk about in great detail in section \ref{Sec:new deconstruction} of this work). Consequently, lower dimensional EFTs containing multiple massive spin-2 fields are a generic prediction of these sorts of models, hence it is important to think about them as a means by which us humans, as 4-dimensional observers, might test for the imprints of extra dimensional physics. Thankfully, over the last decade and a half, our understanding of EFTs of massive gravity has undergone something of a revolution.

The tale of trying to give mass to the graviton in a consistent manner dates back to the time of Fierz and Pauli, who in 1939 first wrote down the linearised theory of a massive, self-interacting spin-2 field \cite{FierzPauli}. As is well-known, general relativity (GR) constitutes the unique nonlinear completion of the linear Fierz-Pauli theory describing a self-interacting, \emph{massless} spin-2 field \cite{Gupta,Weinberg,Deser,Feynman_lecs,GR_from_QG}. However, for nearly a century, it was thought that a similarly healthy nonlinear completion of massive gravity was impossible, owing to the emergence of the so-called Boulware-Deser (BD) ghost -- a problematic scalar mode equipped with wrong-sign kinetic term, signalling an instability of the vacuum -- once nonlinear interactions were taken into account \cite{BD_Ghosts,BD_ghost_explicit}. Because of this, there was little work on massive gravity throughout the 20th century, as the theory was thought to be pathological. However, it turned out that the original BD analysis did not consider all possible interaction terms, and a breakthrough came much later, in 2010, when a viable nonlinear theory of massive gravity was constructed \cite{dRGT_1,dRGT_2,first_action_metric_form} and subsequently proved to be free of the BD ghost \cite{ghost_freedom_flat_ref,ghost_freedom_general_ref,ghost_freedom_stuckelberg,hamiltonian_analysis,Kluson,Kluson_note,Covariant_approach_no_ghosts}. The theory, built upon groundwork laid earlier in \cite{EFTforMGs,Creminelli}, goes by the name of dRGT massive gravity, after its progenitors: de Rham, Gabadadze and Tolley (there were important contributions also by Hassan and Rosen \cite{first_action_metric_form,ghost_freedom_flat_ref,ghost_freedom_general_ref,ghost_freedom_stuckelberg}). It gives a mass to the graviton via a framework in which the physical spacetime metric interacts with some auxiliary reference metric that one inserts by hand, in a special manner that exorcises the ghost. Typically, the reference metric is taken to be Minkowski, although one is free to be more general if one so wishes. By providing a kinetic term for the reference metric, thereby promoting it to a second dynamical field, one obtains the theory of ghost free bigravity \cite{HR1,HR2}. The generalisation to multiple interacting metric fields followed soon after in \cite{interacting_spin2}, although this general theory is only ghost free up to certain conditions, upon which we shall elaborate in section \ref{Sec:review}. 

These theories of multiple interacting spin-2 fields are all encompassed by the umbrella term \emph{`multi-gravity'}, thanks to their construction in terms of multiple interacting metric tensor fields. For further details regarding multi-gravity's development and phenomenology, we refer the reader to the excellent and comprehensive reviews \cite{dR_review,Hinterbichler_review} on massive gravity, as well as \cite{bigravity_review} on bigravity.

Understanding how multi-gravity arises from a more standard gravitational theory (i.e. with only a single massless graviton) in higher dimensions, in the spirit of Kaluza-Klein, is an example of the procedure of \emph{dimensional deconstruction}; the idea is to simply take the higher dimensional gravitational theory and then discretise the extra dimension(s) on a lattice with a finite number of sites. Deconstruction actually has its origins in non-gravitational gauge theories \cite{deconstruction_gauge_theory,KK_deconstruction,EW_sym_breaking_deconstruction}, but the basic procedure applies equally well to gravity too \cite{EFTforMGs,Schwartz_construction,discrete_grav_dims,multigrav_from_ED,Deffayet_deconstruction_review,deconstructing_dims}; it is related to the more standard KK procedure, whereby one integrates out the extra dimensions instead of discretising them, by a discrete Fourier transform \cite{deconstructing_dims}. We shall elaborate fully on how deconstruction works in section \ref{Sec:old deconstruction}, but the starting point is always to consider the $(D+1)$ ADM decomposition of the higher dimensional metric \cite{ADM}, where one foliates the extra dimension by hypersurfaces that are linked together by means of the \emph{lapse} and \emph{shift} fields (see section \ref{Sec:old deconstruction} for definitions of these fields). The induced metrics on these hypersurfaces become analogous to the multi-gravity metrics on the corresponding lattice sites, and the shift vectors become analogous to the Stückelberg fields that restore diffeomorphism invariance to the deconstructed theory (we will introduce the Stückelberg mechanism in section \ref{Sec:stuckelberg}). 

However, the lapse function, which defines the distance between adjacent hypersurfaces, has \emph{no analogue} in standard multi-gravity, because in order to generate the standard ghost free spin-2 interactions upon deconstruction, one must gauge fix the lapse to 1 \emph{before discretising the extra dimension} \cite{deconstructing_dims}; this means it does not appear at all in the standard multi-gravity action, when naively one might expect it to appear as a scalar. This is the cause of a whole host of problems, because the lapse function provides crucial structure to the higher dimensional theory: in particular, it ensures that (in the case where the higher dimensional theory is pure GR) there is diffeomorphism invariance along the extra dimension, and it also enforces the higher dimensional Hamiltonian constraint through its equation of motion. Clearly, these pieces of structure are then missing in the resulting multi-gravity theory and so multi-gravity in its current form cannot genuinely arise from the dimensional deconstruction of higher dimensional GR. In fact, the suggestion from \cite{deconstructing_dims} is that the lack of an analogue for the lapse in standard multi-gravity is the reason why it becomes strongly coupled at such a low energy scale (again, more on this in section \ref{sec:fixing lapse}), and that an improved deconstruction procedure that keeps the lapse free may resolve some or all of these issues. Granted, the resulting lower dimensional theory will not be standard multi-gravity, but it should still be something closely related.

In this work, we develop such a procedure, and show that the (tentatively) correct deconstructed theory is a hybrid of standard multi-gravity coupled to a collection of scalar fields, which correspond to the value of the higher dimensional lapse function on the different hypersurfaces/sites. These scalar fields were not present in standard multi-gravity, and it is precisely their equations of motion that encode the dynamics of the lapse upon taking the continuum limit. Indeed, we explicitly demonstrate, using the concrete 5-dimensional example of Randall-Sundrum brane cosmology (see e.g. \cite{Carsten_review,Langlois_review}), that the field equations and constraints of this new 4-dimensional multi-scalar, multi-gravity theory encode \emph{all} of the field equations, constraints and junction conditions of 5-dimensional GR in the continuum limit. However, we stress that one should still confirm that the new 4-dimensional theory remains valid all the way up to the 5-dimensional Planck scale, a calculation that we save for future work. Nevertheless, our results motivate us to consider the deconstructed theory as an interesting theory of modified gravity in its own right (indeed, even standard multi-gravity is a perfectly valid and interesting lower dimensional EFT, without reference to any deconstruction), as we generalise it to arbitrary dimension and away from the continuum limit. We will show that the new scalar equations of motion change the structure of the simplest solutions of standard multi-gravity (the so-called `proportional solutions' \cite{consistent_spin2,prop_bg_multigrav}), in a manner that reflects what we know to happen in higher dimensional GR.

The structure of the paper, then, is as follows: in section \ref{sec:warmup}, we review standard multi-gravity in arbitrary spacetime dimension, introducing its metric, vielbein and Stückelberg formulations, and we review how deconstruction has been approached previously, including why it fails for standard multi-gravity; in section \ref{Sec:new deconstruction}, we develop our new deconstruction procedure that keeps the lapse free and show that it can recover all of the equations and constraints of higher dimensional GR, in the right circumstances, using the example of Randall-Sundrum brane cosmology; in section \ref{Sec:MMST}, we generalise the resulting deconstructed theory to develop the modified gravity theory that we dub `scalar-tensor multi-gravity', and show that the structure of its simplest solutions differ from standard multi-gravity owing to the new scalar field equations; finally we conclude in section \ref{sec:conclusion}.

We work with natural units $c=\hbar=1$ throughout, and always use a mostly-plus metric signature.

\section{Warm up: deconstructing standard multi-gravity}\label{sec:warmup}

\subsection{Review of standard multi-gravity}\label{Sec:review}

The theory of ghost free multi-gravity, as the name suggests, describes multiple metric tensor fields interacting with one another nonlinearly on the same spacetime manifold; at linear level, it describes the propagation of a single massless spin-2 field together with a finite tower of massive spin-2 fields. The theory is based on the symmetry group $\GC{1}\times\hdots\times\GC{N}$, the direct product of diffeomorphisms associated to each metric (although the interactions actually break this down to just the diagonal subgroup that transforms every metric in the same way), just as GR is symmetric under the one GC associated to the general covariance of the Einstein-Hilbert action. 

Multi-gravity may be formulated using two different approaches: one may work in either the \emph{metric formalism}, where the potential governing the interactions between the various metrics is constructed from those metrics directly, or the \emph{vielbein formalism} (also known as the \emph{tetrad formalism}), where it is instead written in terms of wedge products of the different tetrad 1-forms associated to each of the metrics. Both formalisms are very useful in different situations; for example, the vielbein formalism is essential when one wishes to talk of dimensional deconstruction, and is less restrictive than the metric formalism in a manner we will soon come to, whereas the metric formalism more readily facilitates determination of the structure and perturbations of the multi-gravity field equations. 

Equivalence between the two formalisms is not immediate. Indeed, multi-metric and multi-vielbein theories are actually \emph{not} equivalent in general; their equivalence is intimately tied to the satisfaction of the so-called Deser-van Nieuwenhuizen symmetric vielbein condition, which we introduce in the coming section. However, in all situations we will be considering later on, the metric and vielbein formulations are entirely interchangeable, so we will use both throughout this work, depending on which better lends itself to the problem at hand. We will begin by introducing multi-gravity in the metric formalism, as this is probably the most commonly used, and certainly the most familiar to the massive gravity literature, but we will come on to the vielbein formalism afterwards and show how the two can be related.

\subsubsection{Metric formalism}\label{Sec:metric}

The action, $I$, for multi-gravity living on some $D$-dimensional spacetime manifold, $\mathcal{M}_D$, is expressed in the metric formalism as a sum of $N$ Einstein-Hilbert kinetic terms\footnote{Actually, in $D$ dimensions, any combination of Lovelock invariants \cite{Lovelock,Lovelock_bigrav} beyond the Einstein-Hilbert term can be taken as the kinetic term; the important thing is that there are no derivative couplings between the different metrics, as these generically lead to pathologies \cite{EH_uniqueness_1,EH_uniqueness_2,EH_uniqueness_3}. We stick with standard Einstein-Hilbert here because we are interested in deconstructing GR in one dimension higher.} (one for each metric), along with the ghost free dRGT-style interaction potential that couples the various metrics together, and some action $I_M$ for the collective matter fields coupled to the theory (see \cite{dRGT_1,dRGT_2,first_action_metric_form,metric_multigravity,BHs_multigrav,BH2}):
\begin{align}
    I &= I_K + I_V + I_M\label{MultigravAction}
    \\
    I_K &= \sum_{i=0}^{N-1} \frac{M_i^{D-2}}{2} \int \dd[D]x\, \sqrt{-\det g_{(i)}} R_{(i)}
    \\
    I_V &=  -\sum_{i,j}\int \dd[D]x\, \sqrt{-\det g_{(i)}} \sum_{m=0}^{D} \beta_m^{(i,j)} e_m(S_{i\rightarrow j}) \label{MultigravPotentialMetric} \; .
\end{align}
The multi-metric potential is built by summing up the elementary symmetric polynomials, $e_m$, of the eigenvalues of the characteristic building-block matrices:
\begin{equation}\label{Sij}
    S_{i\rightarrow j} = \sqrt{g_{(i)}^{-1}g_{(j)}} \; ,
\end{equation}
together with some constant coefficients $\beta_m^{(i,j)}=\beta_m^{(j,i)}$ (of mass dimension $D$) that characterise the interactions between $\gi{i}$ and $\gi{j}$. In Eq. \eqref{Sij}, the matrix square root is defined in the sense that $(S^2_{i\rightarrow j})^\mu_{\;\nu}=g^{(i)\mu\lambda}g^{(j)}_{\lambda\nu}$, while the elementary symmetric polynomials are given in terms of the eigenvalues, $\lambda_k$, of $S_{i\rightarrow j}$ as:
\begin{align}
        e_0(\lambda_1,\lambda_2,\hdots,\lambda_{D}) &= 1 \\
        e_1(\lambda_1,\lambda_2,\hdots,\lambda_{D}) &= \sum_{1\leq i \leq D} \lambda_i \\
        \vdots\nonumber \\
        e_k(\lambda_1,\lambda_2,\hdots,\lambda_{D}) &= \sum_{1\leq j_1<j_2<...<j_k\leq D} \lambda_{j_1}\hdots\lambda_{j_k} \\
        \vdots\nonumber \\
        e_{D}(\lambda_1,\lambda_2,\hdots,\lambda_{D}) &= \lambda_1 \lambda_2 \hdots \lambda_{D} \, .
\end{align}
They can also be explicitly constructed iteratively in terms of the trace of $S_{i\rightarrow j}$, starting from $e_0(S)=1$, as:
\begin{equation}\label{sym pols}
    e_m(S)=-\frac{1}{m}\sum_{n=1}^m(-1)^n\Tr(S^n)e_{m-n}(S) \; .
\end{equation}

Lastly, since $S_{i\rightarrow j}=S_{j\rightarrow i}^{-1}$, there is a sense in which these interactions are \emph{oriented}: we say that a term in the potential, Eq. \eqref{MultigravPotentialMetric}, that contains $S_{i\rightarrow j}$ (\emph{not} $S_{j\rightarrow i}$) is positively oriented with respect to the $i$-th metric and negatively oriented with respect to the $j$-th metric. The orientation of an interaction with respect to a given metric affects the form of that metric's field equations, as we will see. It is also simply an artifact of the way one chooses to write down the potential and its interaction coefficients. To see this, note that the following identity holds on the building-blocks of the potential:
\begin{equation}\label{swap_orientation}
    \sqrt{-\det g_{(i)}}\sum_{m=0}^D \beta_m^{(i,j)}e_m(S_{i\rightarrow j}) = \sqrt{-\det g_{(j)}}\sum_{m=0}^D \beta_{D-m}^{(i,j)}e_m(S_{j\rightarrow i}) \; ,
\end{equation}
which shows that one can always consider any given positively oriented interaction as a negatively oriented one simply by redefining the interaction coefficients -- the orientation is nothing mysterious; it is just a manifestation of what one decides to call $\beta_m$ when writing down a particular model.

\begin{figure}[h!]
\centering
    \includegraphics[width=0.6\textwidth]{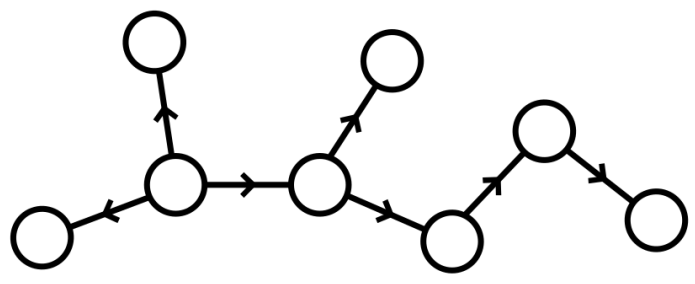}
    \caption{Directed theory graph representing some generic multi-metric model. The circular nodes represent different metrics, the edges indicate interactions and the arrows point in the direction of positive interaction orientation. Each metric generically has a number of interactions of either orientation, and each edge contributes a term to the field equations of the two metrics it connects; these terms are orientation-dependent.}
    \label{fig:interaction structure}
\end{figure}

Given a multi-metric potential, the simplest way to view the corresponding interaction structure is as a directed graph \cite{SC_and_graph_structure,prop_bg_multigrav,BH2}, as depicted in figure \ref{fig:interaction structure}. Symmetries of a particular multi-metric model under permutations of the metric labels and swapping of interaction orientations can then be equivalently viewed as symmetries of said model's directed theory graph.

The multi-metric interactions are not entirely arbitrary: in order to fully exorcise the BD ghost, the interaction structure cannot contain any cycles (a cycle is e.g. $1\rightarrow2\rightarrow3\rightarrow1$, so that the potential is built from \emph{all three} of $S_{1\rightarrow2}$, $S_{2\rightarrow3}$ and $S_{3\rightarrow1}$; in other words, it is a loop in the theory graph) \cite{cycles,ghost_freedom_multigravity}. Precisely, including a cycle leads to the loss of the secondary constraints that would otherwise kill the ghostly degrees of freedom, so those ghosts survive. The same requirement extends also into the matter sector, so that within $I_M$ one is only able to couple entirely \emph{separate} matter sectors to separate metrics -- one can imagine introducing an energy-momentum tensor into the theory graph as a different type of node (say, depicted by a square); again no loops are permitted to form, between either circles, squares or their combinations \cite{ghost_doubly_coupled,on_matter_couplings,ghosts_matter_couplings_rev}\footnote{There is the notable exception, however, where a \emph{single} matter source can be coupled to multiple metrics in a ghost free manner through the special `effective' metric considered in \cite{on_matter_couplings,ghost_freedom_eff_metric,matter_coupling_multigravity,generalised_matter_couplings,eff_matter_coupling_completion}.}.

The field equations that arise from the action \eqref{MultigravAction} are as follows:
\begin{equation}\label{Einstein eqs}
    M^{D-2}_{i} \Gi{i} + \Wi{i} = \Tmunui{i} \; ,
\end{equation}
where the new term $W$ characterises the effect of the interactions over and above the standard GR interactions. In the metric formalism, it is explicitly given by:
\begin{equation}\label{W_metric}
    \Wi{i} = \sum_j \sum_{m=0}^D(-1)^m\beta_m^{(i,j)}Y_{(m)\nu}^\mu(S_{i\rightarrow j})
    +\sum_k\sum_{m=0}^D (-1)^m\beta_{D-m}^{(k,i)}Y_{(m)\nu}^\mu (S_{i\rightarrow k}) \; ,
\end{equation}
where (with respect to the $i$-th metric) $j$ denote positively oriented interactions, $k$ denote negatively oriented interactions, and we define the matrices:
\begin{equation}\label{Y_def}
    Y_{(m)}(S) = \sum_{n=0}^m (-1)^n S^{m-n}e_n(S) \; .
\end{equation}
A useful identity one can show using Eq. \eqref{sym pols} is that:
\begin{equation}\label{Trace Y}
    \Tr Y_{(m)}(S) = (-1)^m (D-m) e_m(S) \; .
\end{equation}

Lastly, owing to the Bianchi identities on the Einstein tensors, as well as the general covariance of the total matter sector, the $W$-tensors are subject to the constraint \cite{ClockworkCosmo}:
\begin{equation}\label{W sum}
    \sum_{i=0}^{N-1} \sqrt{-\det g_{(i)}}\nabla^{(i)}_{\mu}\Wi{i} = 0 \; .
\end{equation}
Whenever matter couples to one site only, or when there is no matter coupling at all, the divergences of each $W$-tensor (i.e. each term in the above sum) must instead vanish \emph{individually}. For all situations we will consider in this work, this will indeed be the case\footnote{For example, coupling to one distinguished metric will become analogous in the continuum limit to placing matter on a brane at a distinguished location in the extra dimension.}, so we will have:
\begin{equation}\label{W constraint}
    \nabla^{(i)}_{\mu}\Wi{i} = 0 \;\;\; \forall \, i \; .
\end{equation}
This condition is referred to as the \emph{Bianchi constraint}; it tells us that there can be no flow of energy-momentum across the interacting metrics.

\begin{figure}[h!]
\centering
    \includegraphics[width=0.3\textwidth]{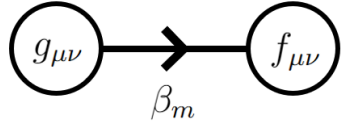}
    \caption{Directed theory graph for bigravity, with $N=2$ metrics. In bigravity, both metrics are treated on the same footing, as one can show using Eq. \eqref{swap_orientation} that the theory is invariant under the simultaneous exchanges of $g\leftrightarrow f$, $M_g\leftrightarrow M_f$ and $\beta_m\leftrightarrow\beta_{D-m}$.}
    \label{fig:bigravity}
\end{figure}

As an instructive example to see how this all works, one may consider the theory of bigravity -- the simplest of the multi-metric theories that contains exactly two metrics, usually denoted $\gi{0}\equiv g_{\mu\nu}$ and $\gi{1}\equiv f_{\mu\nu}$. The theory graph consists of just two nodes adjoined by a single line, as in figure \ref{fig:bigravity}.

The single interaction, parametrised by one set of coefficients $\beta_m^{(g,f)}\equiv\beta_m$, is positively oriented with respect to $g_{\mu\nu}$ and negatively oriented with respect to $f_{\mu\nu}$, so the field equations are:
\begin{align}
    M_g^{D-2}\Gi{g} + \sum_{m=0}^D(-1)^m\beta_m Y_{(m)\nu}^\mu(S_{g\rightarrow f}) &= \Tmunui{g} \; ,
    \\
    M_f^{D-2}\Gi{f} + \sum_{m=0}^D(-1)^m\beta_{D-m} Y_{(m)\nu}^\mu(S_{f\rightarrow g}) &= \Tmunui{f} \; ,
\end{align}
which are precisely the standard bigravity field equations quoted in e.g. \cite{Bigravity_gen_dim,bigravity_review}.

\subsubsection{Vielbein formalism}\label{Sec:vielbein}

In the vielbein formalism, one instead expresses everything given above in the convenient language of differential forms, using the tetrad 1-forms $e^{(i)a}=\vbein{i}{\mu}{a} \dd x^\mu$ in place of the metrics, with the vielbeins defined in the usual way through $\gi{i} = \vbein{i}{\mu}{a} \vbein{i}{\nu}{b} \eta_{ab}$. The action in this language is now written as (see e.g. \cite{interacting_spin2,ClockworkCosmo,BHs_multigrav}):
\begin{align}\label{MultigravActionVbein}
    I &= I_K + I_V + I_M
    \\
    I_K &= \sum_{i=0}^{N-1} \frac{M_i^{D-2}}{2} \int_{\mathcal{M}_D} R^{(i)}_{ab} \wedge \hodge^{(i)} e^{(i)ab}\label{MultigravKinetic}
    \\
    I_V &= -\sum_{i_1\hdots i_D=0}^{N-1} \int_{\mathcal{M}_D} \varepsilon_{a_1\hdots a_D} T_{i_1\hdots i_D} e^{(i_1)a_1}\wedge\hdots\wedge e^{(i_D)a_D}\label{MultigravPotential} \; .
\end{align}
The kinetic term, of course, is just a rewriting of the standard Einstein-Hilbert term for each tetrad in terms of the curvature form, $R^{(i)}_{ab}$, of the $i$-th (Levi-Civita) connection, with the shorthand $e^{(i)ab\hdots}$ meaning $e^{(i)a}\wedge e^{(i)b}\wedge\hdots$, and $\hodge^{(i)}$ being the Hodge star associated to the $i$-th tetrad. 

The potential is now built from the wedge products of the various tetrads, with some new symmetric coefficients $T_{i_1\hdots i_D}=T_{(i_1\hdots i_D)}$ (again of mass dimension $D$) to characterise the interactions. These coefficients are analogous to the $\beta_m^{(i,j)}$ of the metric formalism, but they are not necessarily equivalent; indeed, as alluded to at the start of this section, not all multi-gravity theories described by the multi-vielbein action \eqref{MultigravActionVbein} can be equivalently expressed in the multi-metric language of Eq. \eqref{MultigravAction}. In fact, this happens only when the so-called Deser-van Nieuwenhuisen symmetric vielbein condition:
\begin{equation}\label{SymmetricVierbeinCondition}
    e_{\;\;\;\;\;a}^{(i)\mu}e_\mu^{(j)b} = e^{(i)\mu b}e^{(j)}_{\mu a} \; ,
\end{equation}
is satisfied, which allows one to trade off products of vielbeins for the $S_{i\rightarrow j}$ matrices of the metric formalism (see e.g. \cite{interacting_spin2}). The only known multi-vielbein models that satisfy this condition and hence have a metric description of the form \eqref{MultigravAction} are those involving exclusively pairwise interactions; in terms of the $\Tcoeffs$, this restricts one to only permit terms of the form $T_{iiii\hdots}$, $T_{jiii\hdots}$, $T_{jjii\hdots}$ and so on. 

Provided that there are no cycles in the interaction structure, as we stated in the previous subsection, these pairwise interacting models are the only ghost free theories one may write down in the metric formalism. In general, multi-vielbein theories with arbitrary $\Tcoeffs$ are ghostly, so for a time it was thought that the only ghost free multi-vielbein theories one could write down were those with pairwise interactions that could be equivalently expressed in metric form \cite{vielbein_to_rescue}. However, this turned out not to be true, as a more recent class of multi-vielbein models free of the BD ghost was found in \cite{beyond_pairwise_couplings} (its spectrum was determined in \cite{mass_spectrum_multivielbein}), where the interaction term can be rewritten as the determinant of the sum of the tetrads; in terms of the $\Tcoeffs$, this means one is restricted to only terms that factorise into $\Tcoeffs=T_{i_1}T_{i_2}\hdots T_{i_D}$. This class of models \emph{does} allow for cycles, and does not yet have an equivalent metric formulation, so in this sense multi-vielbein models are more general than multi-metric models.

Still, throughout this work, we will only be considering models with pairwise interactions, since we will see that the continuum limit only makes sense when all metrics/tetrads interact in this way. Therefore, the metric and vielbein formalisms will be interchangeable for us; precisely, the $\Tcoeffs$ of the vielbein formalism, upon restricting oneself to considering only pairwise interactions, are related to the $\beta_m^{(i,j)}$ of the metric formalism as \cite{BHs_multigrav}:
\begin{align}
    D!\,T_{iiii\hdots i} &= \sum_j\beta_0^{(i,j)} + \sum_k\beta_D^{(k,i)}\label{betas1} \\
    D!\,T_{\underbrace{j\hdots j}_m\underbrace{i\hdots i}_{D-m}} &= \beta_m^{(i,j)}\label{betas2} \; ,
\end{align}
where again $j$ and $k$ refer respectively to positively and negatively oriented interactions with respect to $\gi{i}$, so the sense of interaction orientation from the metric formalism is encoded in the vielbein formalism within the structure of $T_{iiii\hdots i}$.

The field equations are as in \eqref{Einstein eqs}, and the $W$-tensor components in vielbein form are given by (see the appendix of \cite{BHs_multigrav} for the derivation, and differential form version of this expression):
\begin{equation}\label{W tensor vielbein}
    \Wi{i} = D!\vbein{i}{\nu}{a}\ivbein{i}{\mu}{[a}\ivbein{i}{\lambda_1}{b_1}\hdots\ivbein{i}{\lambda_{D-1}}{b_{D-1}]} \sum_{j_1\hdots j_{D-1}} \mathcal{P}(i) T_{ij_1\hdots j_{D-1}} \vbein{j_1}{\lambda_1}{b_1}\hdots\vbein{j_{D-1}}{\lambda_{D-1}}{b_{D-1}} \; ,
\end{equation}
with $\mathcal{P}(i)$ counting the number of times the index $i$ appears in the interaction coefficients i.e. a term with $T_{ij_1\hdots j_{D-1}}$ has $\mathcal{P}(i)=1$, a term with $T_{iij_2\hdots j_{D-1}}$ has $\mathcal{P}(i)=2$, and so on. One may show that whenever one takes exclusively pairwise interactions this is equivalent to Eq. \eqref{W_metric} in the metric formalism, by substituting in Eqs. \eqref{betas1} and \eqref{betas2} for the $\Tcoeffs$, then contracting any vielbeins with the same index into generalised Kronecker deltas. After this contraction, one is left with terms involving powers of $\ivbein{i}{\mu}{a}\vbein{j}{\mu}{a}$, which is precisely $S_{i\rightarrow j}$ given in terms of the vielbeins associated to $\gi{i}$ and $\gi{j}$.

\subsubsection{Restoring diffeomorphism invariance}\label{Sec:stuckelberg}

As mentioned earlier, multi-gravity in either formalism is based on the symmetry group $\GC{1}\times\hdots\times\GC{N}$, the direct product of $N$ diffeomorphisms generated by $N$ vectors $\xi_i$ that transform the metrics separately as $\delta\gi{i} = \mathcal{L}_{\xi_i}\gi{i}$. The non-interacting theory is invariant under this entire collection of diffeomorphisms; turning on the interactions, the symmetry is broken to the single surviving diagonal subgroup that transforms all of the metrics simultaneously by the same vector $\xi$. It is easy to see that this is the only way to keep $S_{i\rightarrow j}$ invariant; the transformation on $g_{(i)}^{-1}$ undoes the transformation on $g_{(j)}$ only when the metrics each transform by the same amount. The theory consequently propagates a single massless spin-2 field that is invariant under transformations of this subgroup, together with $N-1$ massive spin-2 fields, all of which are combinations of the original metric perturbations -- see \cite{BHs_multigrav,BH2} for the complete linearised theory.

However, as we know, diffeomorphism invariance is a \emph{gauge} symmetry. It is not a real physical symmetry of a system; rather, it is a redundancy of description: any theory that appears diffeomorphism invariant may be written in a manner that is not manifestly so simply by choosing a preferred coordinate system. Similarly, any theory with manifestly broken diffeomorphisms may have them restored via the introduction of new gauge degrees of freedom that account for the symmetry breaking -- this is the essence of the \emph{Stückelberg} trick. In multi-gravity, it is easiest to demonstrate how this works in the metric formalism \cite{interacting_spin2_stuckelberg}, following the pioneering work of Arkani-Hamed, Georgi and Schwartz back in 2003 \cite{EFTforMGs}.

The starting point is to note that each $\GC{i}$ factor sitting within the direct product $\GC{1}\times\hdots\times\GC{N}$ acts in the same way as would a change of coordinates\footnote{Of course, transformations under each individual $\GC{i}$ factor are not really coordinate transformations: since every metric lives on the same spacetime, a genuine coordinate transformation would transform every metric in the same way -- coordinate transformations actually generate the diagonal subgroup!} from ${x^\mu \rightarrow f^\mu_i(x)}$, transforming the distinguished metric $\gi{i}$ as:
\begin{equation}\label{diffeo}
    \gi{i}(x) \rightarrow \partial_\mu f_i^\alpha \partial_\nu f_i^\beta g^{(i)}_{\alpha\beta}(f_i(x)) \; .
\end{equation}
The vectors $\xi_i$ mentioned above are the infinitesimal generators of these diffeomorphisms; that is, to lowest order, $f_i^\mu(x) = x^\mu- \xi_i^\mu$. The infinitesimal version of Eq. \eqref{diffeo} then becomes $\delta\gi{i} = \mathcal{L}_{\xi_i}\gi{i}$, as we wrote earlier.

Considering the metric tensor itself i.e. $\mathbf{g}=g_{\mu\nu}(x)\dd x^\mu \otimes \dd x^\nu$, rather than just its components $g_{\mu\nu}$, one may express Eq. \eqref{diffeo} as a functional composition \cite{EFTforMGs}:
\begin{equation}\label{func_decomp}
    \mathbf{g}_{(i)} \xrightarrow{f_i} \mathbf{g}_{(i)} \circ f_i \; ,
\end{equation}
where the coordinate basis 1-forms transform as $\dd x^\mu \rightarrow \dd f_i^\mu = \partial_\alpha f_i^\mu \dd x^\alpha$. 

Each metric $\mathbf{g}_{(i)}$ lies in a tensor representation of $\GC{i}$ \emph{only}; they are scalars as far as the other $\GC{j}$ factors that exist within the product $\GC{1}\times\hdots\times\GC{N}$ are concerned. Therefore, only transformations wrought by $f_i^\mu$ alter the metrics in the manner defined by Eq. \eqref{func_decomp}; under action of the other $\GC{j}$, the metrics are unchanged:
\begin{equation}
    \mathbf{g}_{(i)} \xrightarrow{f_j} \mathbf{g}_{(i)} \;\;\;\;\;\; (i\neq j) \; .
\end{equation}
Thus, under a gauge transformation of the entire direct product group of diffeomorphisms, $\GC{1}\times\hdots\times\GC{N}$, generated by all the functions $f_1^\mu,\hdots,f_N^\mu$ simultaneously, each metric $\mathbf{g}_{(i)}$ only sees the part coming from its corresponding $f_i^\mu$.

The goal now is to introduce new gauge degrees of freedom, called Stückelberg fields, that allow us to modify the multi-gravity potential in such a way that the theory becomes manifestly invariant under the entire product $\GC{1}\times\hdots\times\GC{N}$. In practice, this means making all of the $S_{i\rightarrow j}$ invariant under the various sub-products $\GC{i}\times\GC{j}$, since each of the kinetic terms depends on one metric only and is hence already manifestly diffeomorphism invariant. Upon fixing a particular gauge for the Stückelberg fields, one should still be able to recover the original action where these symmetries appear broken (aside from the diagonal subgroup), and where these new gauge degrees of freedom have been eaten to give all but one of the spin-2 fields masses. Let us see how this procedure works.

Firstly, remember that $S_{i\rightarrow j}=\sqrt{g_{(i)}^{-1}g_{(j)}}$, and that $\mathbf{g}_{(i)}$ and $\mathbf{g}_{(j)}$ lie in tensor representations of $\GC{i}$ and $\GC{j}$, respectively, while they are scalars under the other. To make $S_{i\rightarrow j}$ invariant under $\GC{i}\times\GC{j}$, we need to either replace $\mathbf{g}_{(i)}$ with an object that transforms as a tensor under only $\GC{j}$, or $\mathbf{g}_{(j)}$ with an object that transforms as a tensor under only $\GC{i}$. It doesn't matter which way one decides to do this, as they are ultimately related by dualities \cite{interacting_spin2_stuckelberg,decoupling_limit_multigravity} -- we will choose the latter option for concreteness. The desired objects, which we denote $\mathbf{\tilde{g}}_{(i)}$, are built as follows:
\begin{equation}\label{stuckelberg gs}
    \mathbf{\tilde{g}}_{(i)} = \mathbf{g}_{(j)} \circ Y_{(j,i)} \; ,
\end{equation}
or in components:
\begin{equation}
    \tilde{g}^{(i)}_{\mu\nu}(x) = \partial_\mu Y_{(j,i)}^\alpha \partial_\nu Y_{(j,i)}^\beta g^{(j)}_{\alpha\beta}(Y_{(j,i)}(x)) \; .
\end{equation}
The $Y_{(j,i)}^\mu(x)$ are our newly introduced Stückelberg fields, which are charged under \emph{both} factors of the product $\GC{i}\times\GC{j}$ (they lie in the tensor product representation of the fundamental of $\GC{i}$ with the anti-fundamental of $\GC{j}$) to transform as:
\begin{equation}
    Y_{(j,i)} \xrightarrow{f_i, \;f_j} f_j^{-1} \circ Y_{(j,i)} \circ f_i \; ,
\end{equation}
which again in components reads:
\begin{equation}
    Y_{(j,i)}^\mu(x) = (f_j^{-1})^\mu Y_{(j,i)}(f_i(x)) \; ,
\end{equation}
so the Stückelberg fields $Y_{(j,i)}^\mu(x)$ can be thought of as defining the pullback map of the metric $\mathbf{g}_{(j)}$ from site $j$ to site $i$. In terms of the graph structure of section \ref{Sec:metric}, the different $Y_{(j,i)}$ fields live on the links between different nodes; for a theory with $N$ metrics, there are $N-1$ such links, hence there are $N-1$ sets of Stückelberg fields. In this sense, the theory graphs we defined earlier are really \emph{quiver diagrams} \cite{Moose,Quivers,deconstruction_gauge_theory} for the gravitational gauge theory whose gauge group is the direct product of diffeomorphisms, $\GC{1}\times\hdots\times\GC{N}$; in quiver language, the Stückelberg fields are the link fields.

Under a gauge transformation, our new objects $\mathbf{\tilde{g}}_{(i)}$ consequently transform as:
\begin{equation}
    \mathbf{\tilde{g}}_{(i)} \xrightarrow{f_i, \; f_j} \mathbf{g}_{(j)} \circ f_j \circ f_j^{-1} \circ Y_{(j,i)} \circ f_i = \mathbf{\tilde{g}}_{(i)} \circ f_i \; ,
\end{equation}
in other words, they transform as tensors under $\GC{i}$ only, and as scalars under $\GC{j}$, which is exactly what we wanted! Hence, one may replace in the multi-gravity action all of the $S_{i\rightarrow j}$ matrices with the new building-block matrices $\tilde{S}_{i\rightarrow j}$ defined by:
\begin{equation}\label{Stuckelberg metric}
    S_{i\rightarrow j} \rightarrow \tilde{S}_{i\rightarrow j} \equiv \sqrt{g_{(i)}^{-1} \tilde{g}_{(i)}} = \sqrt{g^{(i)\mu\lambda} \partial_\lambda Y_{(j,i)}^\alpha \partial_\nu Y_{(j,i)}^\beta g^{(j)}_{\alpha\beta}} \; ,
\end{equation}
which are each invariant under their corresponding sub-product $\GC{i}\times\GC{j}$\footnote{Precisely, both $g_{(i)}^{-1}$ and $\tilde{g}_{(i)}$ transform as tensors only under the $\GC{i}$ factor, and they transform by complementary amounts that cancel one another (since one of them is inverted); both transform as scalars under the $\GC{j}$ factor, so $\tilde{S}_{i\rightarrow j}$ is invariant under the product $\GC{i}\times\GC{j}$.}; then, the full interaction term involving sums over all the different $\tilde{S}_{i\rightarrow j}$ will be invariant under the full product $\GC{1}\times\hdots\times\GC{N}$. By including $N-1$ sets of Stückelberg fields to construct the interaction terms in this way, one restores the $N-1$ broken diffeomorphism invariances to the theory.

It is important to note that one neither gains nor loses any physical information by writing the theory in terms of the Stückelberg fields -- they are just gauge fields after all, so one can always choose to fix them to the so-called \emph{unitary gauge}, where all $Y_{(j,i)}^\mu = x^\mu$. In unitary gauge, $\tilde{S}_{i\rightarrow j}=S_{i\rightarrow j}$, so one recovers the standard form of the multi-gravity action \eqref{MultigravAction}, invariant under only the diagonal subgroup of diffeomorphisms. Indeed, one may show that the equations of motion for the Stückelberg fields, if one chooses to include them explicitly, are completely equivalent to the Bianchi constraint in the unitary gauge theory \cite{ghost_freedom_stuckelberg}. Precisely, one finds that:
\begin{equation}\label{stuckelberg bianchi}
    \nabla^{(i)}_\mu \Wi{i} = \sum_j \frac{\delta I}{\delta Y_{(j,i)}^\mu} \partial_\nu Y_{(j,i)}^\mu - \sum_k \frac{\delta I}{\delta Y_{(i,k)}^\mu} \partial_\nu Y_{(i,k)}^\mu\; ,
\end{equation}
so the divergences of the $W$-tensors vanish provided that the Stückelberg fields satisfy their Euler-Lagrange equations.

The utility of including the Stückelberg fields lies in the understanding that they offer. In a general situation, one may expand the various $Y_{(j,i)}^\mu$ around the identity as:
\begin{equation}\label{Goldstones}
    Y_{(j,i)}^\mu(x) = x^\mu + \pi^\mu_{(j,i)}(x) \; ,
\end{equation}
then the $\pi^\mu_{(i,j)}$ are essentially the Goldstone bosons associated to the broken diffeomorphisms, which are eaten by the interactions in unitary gauge to give the spin-2 fields their masses. One may decompose these Goldstone modes into their transverse vector and longitudinal scalar components as:
\begin{equation}
    \pi_{(j,i)}^\mu = g^{(i)\mu\nu} \left(A^{(j,i)}_{\nu} + \partial_\nu \phi^{(j,i)}\right) \; ;
\end{equation}
the Goldstone boson equivalence theorem then ensures that, in the `decoupling limit', which we will introduce more concretely in section \ref{sec:fixing lapse}, $A^{(j,i)}_\mu$ and $\phi^{(j,i)}$ become the helicity-1 and helicity-0 modes of the various massive spin-2 fields, respectively \cite{EFTforMGs}. Many of the fiddly and occasionally problematic aspects of massive gravity theories are related to the behaviour of the helicity-0 modes (e.g. the vDVZ discontinuity \cite{vD,Zakharov}, the Vainshtein mechanism \cite{Vainshtein,Recovering_GR,Vainshtein_decoupling_limit,Vainshtein_recovery,Vainshtein_review} etc. -- see the reviews \cite{Hinterbichler_review,dR_review} for details), so it is often advantageous to keep track of them explicitly by introducing the Stückelberg fields. For us, we will see shortly that the Stückelberg fields have the natural interpretation under dimensional deconstruction as encoding the shift vector in the extra dimensional theory.

We note lastly that one may also perform the Stückelberg trick in the vielbein formalism, if one so wishes; the only difference is that the vielbein action \eqref{MultigravActionVbein} also has additional broken local Lorentz invariances (associated to the Latin indices) that must be Stückelberged. Precisely, one can imagine constructing and rewriting the interactions in terms of the new objects \cite{complete_decoupling_limit,deconstructing_dims}:
\begin{equation}\label{vielbein Stuckelberg}
    \tilde{e}^{(i)a}_{\mu}(x) = \partial_\mu Y_{(j,i)}^\alpha \Phi^a_{(j,i)b} \vbein{j}{\alpha}{b}(Y_{(j,i)}(x)) \; ,
\end{equation}
where $\Phi^a_{(i,j)b}$ are the Stückelberg fields for the broken Lorentz invariances. Note that the manner in which the $\Phi^a_{(j,i)b}$ are introduced mimics the way a standard local Lorentz transformation acts on the vielbeins, taking $\vbein{j}{\mu}{a}\rightarrow \Lambda^a_{(j)b}\vbein{j}{\mu}{b}$, just as the $Y_{(i,j)}^\mu$ fields are introduced in a manner that mimics the way diffeomorphisms act on the spacetime indices. Replacing all instances of $\vbein{j}{\mu}{a}$ by $\tilde{e}^{(i)a}_\mu$ in the multi-vielbein potential results in a theory invariant under the direct products of both diffeomorphisms and local Lorentz invariances (at the cost of introducing these new gauge degrees of freedom). In theories without interaction cycles, the equations of motion for the Lorentz Stückelberg fields enforce the Deser-van Nieuwenhuisen symmetric vielbein condition \eqref{SymmetricVierbeinCondition} that ensures equivalence between the metric and vielbein formulations of multi-gravity \cite{complete_decoupling_limit}.

\subsection{The continuum limit and resulting extra dimensional theory}\label{Sec:old deconstruction}

\begin{figure}[h!]
    \centering
    \begin{subfigure}{0.75\textwidth}
        \includegraphics[width=\textwidth]{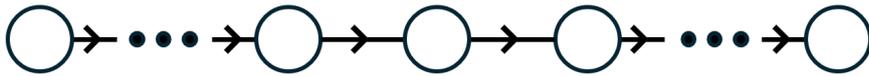}
        \caption{Chain-type interaction}
        \label{chain}
    \end{subfigure}
    \begin{subfigure}{0.34\textwidth}
        \includegraphics[width=\textwidth]{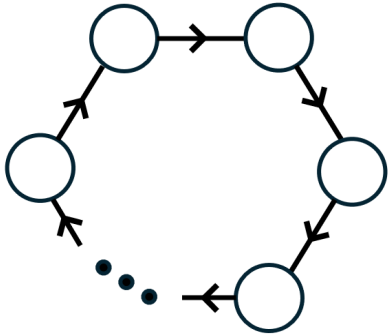}
        \caption{Ring-type interaction}
        \label{ring}
    \end{subfigure}
    \caption{Quiver diagrams/theory graphs corresponding to theories that are amenable to dimensional deconstruction. In the case of the chain, the extra dimension lives on an interval and is bounded at both ends, whereas in the case of the ring, the extra dimension is periodic and lives on the circle, $S_1$.}
    \label{fig:quivers}
\end{figure}

We are at last in a position where we can begin to understand what it means to attempt to take the continuum limit of multi-gravity, but first we must begin with some history. As mentioned in the introduction, dimensional deconstruction is an old idea that has its origin in non-gravitational quiver gauge theories \cite{deconstruction_gauge_theory,KK_deconstruction,EW_sym_breaking_deconstruction}, but the concept is the same for gravitational theories too and the basic starting point is now familiar: one always considers some theory based on a direct product gauge group $\mathcal{S}_1\times\hdots\times\mathcal{S}_N$, containing $N$ fields/sites of a given type, each of which is charged under one of the individual factors $\mathcal{S}_i$ that exist within the product. Originally, the setup involved a collection of scalars charged under different copies of $U(1)$, but in our case of gravitational deconstruction, as we have seen, the fields in question are metrics and the symmetries are diffeomorphisms. One considers the particular scenario where the corresponding quiver diagram (theory graph, in our terms) forms a chain or ring; that is, the $i$-th field interacts only with its nearest neighbours and all the interactions are oriented positively from $i$ to $i+1$, as in figure \ref{fig:quivers}.

One then introduces, through the interaction coefficients that couple the fields, a notion of spacing between the different sites, and upon taking the simultaneous limit where the number of fields is sent to infinity and the lattice spacing is sent to zero (while keeping their product fixed), one arrives at a gauge theory based on a single copy of the underlying symmetry group $\mathcal{S}$ but in one dimension higher. This is called taking the \emph{continuum limit} of the quiver gauge theory. In the gravitational case that we would like to consider, where the symmetry group of the $D$-dimensional multi-gravity theory is $\GC{1}\times\hdots\times\GC{N}$, the continuum theory is then based on a single copy of general covariance in $(D+1)$-dimensional spacetime. Therefore, the lattice spacing in this case has the genuine physical interpretation of representing the separation of codimension-1 spatial hypersurfaces foliating an additional compact spatial dimension, with the various metrics $\gi{i}$ corresponding to the induced metrics on each of these hypersurfaces. If the continuum limit exists, it thus provides us with a powerful way of thinking about both multi-gravity theories with these special types of interaction structure and theories of standard gravity in higher dimensions. 

We note that, in the multi-gravity case, ring-type interactions in the deconstructed theory are ghostly as they contain a cycle. However, since there are no ghosts in the extra dimensional theory (standard higher dimensional gravity, plus maybe a Horndeski scalar \cite{Deconstructing_CW} -- we will see this later on), presumably the masses of the ghosts should be proportional to the inverse lattice spacing, so that they blow up and decouple in the continuum limit\footnote{As far as I am aware, nobody has showed this explicitly yet, but it would be interesting to confirm.}. Regardless, we wish to work with healthy deconstructed theories away from the continuum limit, so we shall henceforth work with chain-type interactions to do our deconstructing. For concreteness, we will also choose to work between multi-gravity in $D=4$ dimensions and standard gravity in $D+1=5$ dimensions, although the process of dimensional deconstruction outlined above functions in qualitatively the same way between any $D$ and $(D+1)$.

Taking chain-type interactions means that our extra dimension lives on an interval, which we parametrise using a new coordinate $y\in [0,L]$, where $L$ is the size of the compact dimension, and the 5-dimensional manifold factorises as:
\begin{equation}
    \mathcal{M}_5 = \mathcal{M}_4 \times [0,L] \; .
\end{equation}
The manifold $\mathcal{M}_5$ thus possesses a 4-dimensional boundary, $\partial\mathcal{M}_5$, with two components: one at $y=0$, which is negatively oriented, and one at $y=L$, which is positively oriented\footnote{Orientation in this sense means that integration over the boundary component in question comes equipped with the appropriate sign; it should not be confused with the multi-gravity interaction orientation of section \ref{Sec:metric}, although the two actually do turn out to be related, as we will see when we come to section \ref{Sec:new deconstruction} -- and explicitly in appendix \ref{App:brane cosmo}.}. Usually when one talks of 5-dimensional gravity, the extra dimension is orbifolded on $S_1/\mathbb{Z}_2$ (the circle that has its top and bottom halves identified), with fixed points of $\mathbb{Z}_2$ at $y=0$ and $y=L$ rather than true boundaries, the $\mathbb{Z}_2$ symmetry helping to ease calculations at these special points \cite{RS1,RS2,Langlois_review,Carsten_review}. In the multi-gravity case, this would correspond to the ring-type interaction in figure \ref{ring}, but with the interaction orientation reversed along the bottom half of the theory graph, and each metric along the top half identified with its corresponding metric along the bottom half. Such a theory is completely equivalent to a chain theory with all contributions of the bulk metrics doubled, so we may as well just consider the chain/interval in the first place.

\begin{figure}[h!]
\centering
    \includegraphics[width=0.68\textwidth]{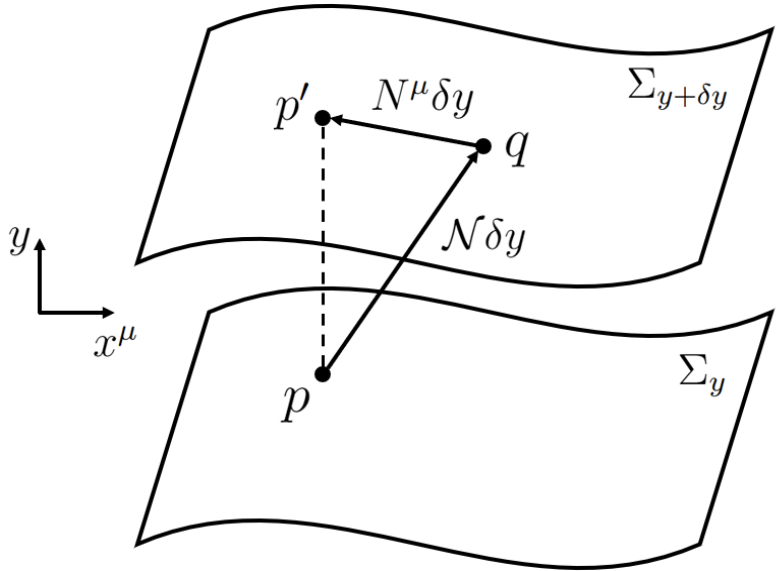}
    \caption{Diagram depicting the spatial ADM decomposition given by Eq. \eqref{ADM}. The hypersurfaces $\Sigma_y$ and $\Sigma_{y+\delta y}$ are a distance $\mathcal{N}\dy$ apart, and the dashed line tracks the evolution (in $y$) of a point $p\in\Sigma_y$ with coordinates $x^\mu$ to the corresponding point $p'\in\Sigma_{y+\delta y}$ with the same coordinates. The normal to $p$ generically points towards a different point $q\in\Sigma_{y+\delta y}$ whose coordinates differ from those of $p'$ by an amount $N^\mu\dy$.}
    \label{fig:ADM}
\end{figure}

Since the different $\gi{i}$ in the deconstructed theory correspond to the induced metrics on different constant-$y$ hypersurfaces in the extra dimension, it is sensible to decompose the 5-dimensional metric using a 4+1 spatial ADM split\footnote{In vielbein form, this implies $E^a = e^a + N^a\dd y$ and $E^5=\mathcal{N}\dd y$, where $E^A$ are the 5-dimensional tetrads defined implicitly in terms of the 5-dimensional metric, $G_{MN}$, via $G_{MN}=e^{\;A}_M e^{\;B}_N \eta_{AB}$.} \cite{ADM,multigrav_from_ED,Deffayet_deconstruction_review,solutions_in_multigrav_deffayet,deconstructing_dims}:
\begin{equation}\label{ADM}
    \dd s^2 = G_{MN}\dd x^M \dd x^N = g_{\mu\nu}\left(\dd x^\mu + N^\mu \dd y\right)\left(\dd x^\nu + N^\nu \dd y\right) + \mathcal{N}^2\dd y^2 \; . 
\end{equation}
The ADM splitting is the most general way to encode the foliation of the 5-dimensional manifold $\mathcal{M}_5$ by 4-dimensional hypersurfaces of constant $y$, which we denote $\Sigma_y$. Each such hypersurface has an induced 4-dimensional metric $g_{\mu\nu}$; the distance between two infinitesimally close hypersurfaces $\Sigma_y$ and $\Sigma_{y+\delta y}$ is $\mathcal{N}\delta y$, which defines the \emph{lapse} field, $\mathcal{N}$; and the normal to a point $p\in\Sigma_y$ with coordinates $x^\mu$ hits $\Sigma_{y+\delta y}$ at a point $q$ with coordinates that in general differ from $x^\mu$ by $N^\mu\delta y$, which defines the \emph{shift} field, $N^\mu$. Figure \ref{fig:ADM} shows the situation more clearly.

As alluded to, one establishes the connection to multi-gravity by discretising the continuous coordinate $y$ into a discrete set of $N$ points, $y_i$, with $i$ running from 0 to $N-1$, separated by distance $\dy$ i.e. $y\rightarrow y_i = i\dy$. The continuum limit we would like to take is $N\rightarrow\infty$ and $\dy\rightarrow0$ with $N\dy=L$ fixed. In this limit, the metric on a given site of the theory graph is simply the induced metric on the hypersurface $\Sigma_{y_i}$:
\begin{equation}\label{discrete metrics}
    \gi{i}(x) = g_{\mu\nu}(x, y_i) \; ,
\end{equation}
while the lapse and shift are also discretised similarly to become a collection of scalar and vector fields living at their corresponding sites:
\begin{align}
    \mathcal{N}_i(x) &= \mathcal{N}(x,y_i) \; ,\label{discrete lapses}
    \\
    N^\mu_i(x) &= N^\mu(x,y_i) \; .\label{discrete shifts}
\end{align}
We have seen that the way the lapse and shift enter the extra dimensional theory is as a means of linking different hypersurfaces together; dealing with this properly in our deconstruction procedure is fiddlier than simply talking about individual lapses or shifts living at particular sites, and requires one to talk instead about quantities that live on the links between different sites in our theory graphs. We will have a lot more to say about this in section \ref{Sec:new deconstruction}; for now though, let us continue on to discuss how this has been approached previously.

\subsubsection{`Top-down' approach}\label{sec:top down}

Throughout the discussion above, no mention was made of any particular 5-dimensional theory, nor was any particular action specified; we only stated that the theory should be gravitational in nature and based on the symmetry group comprising 5-dimensional diffeomorphisms. In the original approaches to the deconstruction of gravity \cite{discrete_grav_dims,Schwartz_construction,multigrav_from_ED,Deffayet_deconstruction_review,solutions_in_multigrav_deffayet}, the goal was to determine the form of the specific multi-gravity theory that becomes 5-dimensional GR in its continuum limit (if such a limit exists). Thus, these approaches all started off with the Einstein-Hilbert action in 5-dimensions and tried to discretise it in a sensible way that leads to a well-defined multi-gravity theory in 4-dimensions; this is why we refer to the approach as `top-down', as it assumes a particular 5-dimensional theory and works backwards from there.

The 5-dimensional Einstein-Hilbert action in terms of our 4+1 ADM variables is given by:
\begin{align}
    I_{(5)} &= \int_{\mathcal{M}} \dd[5]x\sqrt{-G} \left(\frac{M_{(5)}^3}{2} R_{(5)} - 2\Lambda_5\right) + \int_{\partial\mathcal{M}} M_{(5)}^3 K
    \\
    &= \int_{\mathcal{M}} \dd[4]x\dd y\sqrt{-g} \mathcal{N}\left[\frac{M_{(5)}^3}{2} \left(R_{(4)} + K^2 - K_{\mu\nu}K^{\mu\nu}\right) - 2\Lambda_5 \right] + \int_{\partial\mathcal{M}} M_{(5)}^3 K \; ,\label{5D GR}
\end{align}
where $K_{\mu\nu}$ are the components of the extrinsic curvature of the hypersurfaces $\Sigma_y$, defined for any given hypersurface as (half) the Lie derivative of the induced metric on $\Sigma_y$ along its normal vector:
\begin{equation}
    K_{\mu\nu}=\frac12\mathcal{L}_n g_{\mu\nu} \; ,
\end{equation}
and the second integral is the Gibbons-Hawking-York boundary term required to have a well-posed variational problem when the manifold has a boundary \cite{GH,York,York2}.

With the ADM decomposition of Eq. \eqref{ADM}, the coordinate basis form of the normal vector to $\Sigma_y$ is $n = (\partial_y - N^\mu\partial_\mu)/\mathcal{N}$, for which the extrinsic curvature components are explicitly:
\begin{equation}\label{extrinsic curvature}
    K_{\mu\nu} = \frac{1}{2\mathcal{N}}\left(g_{\mu\nu}^{\prime} - D_\mu N_\nu - D_\nu N_\mu \right) \; ,
\end{equation}
with the prime denoting $\partial_y$, and $D_\mu$ being the induced 4-dimensional covariant derivative on $\Sigma_y$\footnote{At a particular site $y_i$, one has $D_\mu=\nabla^{(i)}_{\mu}$, the $i$-th covariant derivative in the multi-gravity theory.}. The authors of \cite{multigrav_from_ED,Deffayet_deconstruction_review,solutions_in_multigrav_deffayet,deconstructing_dims} call the operation of the derivative parts of the normal vector on a function the ``covariant $y$-derivative'', and denote it by $D_y$ i.e.
\begin{equation}
    D_y \equiv \partial_y - N^\mu\partial_\mu \; , \;\;\; K_{\mu\nu} = \frac{1}{2\mathcal{N}}\mathcal{L}_{D_y}g_{\mu\nu} \; .
\end{equation}
We note that in pure GR, the lapse field, $\mathcal{N}$, and shift vector, $N^\mu$, are Lagrange multipliers that enforce the Hamiltonian and momentum constraints, respectively; they are not dynamical and hence can be gauge fixed as one wishes, at least in the continuum theory on its own without reference to any deconstruction\footnote{Remember, the whole \emph{raison d'être} of this paper is that fixing the lapse first and deconstructing the theory second causes problems, which we will soon see.}.

The next step in deconstruction requires that one develops a procedure to discretise the Lie derivatives, replacing them with appropriate finite difference expressions linking neighbouring sites. This is done by introducing to the continuum theory the Wilson line operators defined by \cite{multigrav_from_ED,Deffayet_deconstruction_review}:
\begin{equation}
    W(y', y) = \mathcal{P} \exp \int_y^{y'} \dd \tilde{y} \,\mathcal{L}_{D_{\tilde{y}}} \; .
\end{equation}
These are path-ordered exponentials of Lie derivatives along the normal vectors to $\Sigma_y$, which define mappings from $\Sigma_y$ to $\Sigma_{y'}$. Precisely, one finds that when acting on scalar fields living on $\Sigma_y$, the operation of these Wilson lines takes the explicit form:
\begin{equation}
    W(y', y) = \mathbbm{1} + \int_y^{y'} \dd \tilde{y} \,N^\mu(\tilde{y})\partial_\mu + \int_y^{y'}\dd \tilde{y} \,N^\mu(\tilde{y}) \partial_\mu \int_y^{\tilde{y}} \dd\tilde{y}_1 \,N^{\mu_1}(\tilde{y}_1) \partial_{\mu_1} + \hdots \; ,
\end{equation}
which can be trivially extended to tensor fields of any rank through action on their (scalar) components, since the Leibniz rule for Lie derivatives implies:
\begin{equation}
    W(y', y)(A \otimes B) = W(y',y)A \otimes W(y',y)B \; ,
\end{equation}
for arbitrary tensor fields $A$ and $B$ living on $\Sigma_y$. 

This business of mapping fields from one hypersurface to the next smells very much like the operation of the Stückelberg fields that we introduced in section \ref{Sec:stuckelberg}; indeed, by denoting the action of the Wilson lines on the coordinate functions $x^\mu$ as:
\begin{align}\label{continuum stuckelberg}
    Y^\mu(y', y) &\equiv W(y', y)x^\mu
    \\
    &= x^\mu + \int_y^{y'}\dd\tilde{y}\, N^\mu + \int_y^{y'}\dd \tilde{y} \,N^\nu \int_y^{\tilde{y}} \dd\tilde{y}_1\,\partial_\nu  N^{\mu} + \hdots \; ,
\end{align}
one may express the action of the Wilson lines on any tensor field living on $\Sigma_y$ in the suggestive manner (c.f. Eq. \eqref{stuckelberg gs}):
\begin{equation}
    T_{y'} = W(y',y)T_y = T_y \circ Y(y',y) \; .
\end{equation}
We are almost there, but first note that when $y$ and $y'$ are infintesimally close, one has $Y^\mu = x^\mu + N^\mu\dy + \mathcal{O}(\dy^2)$, showing that the Wilson lines enact the map from $p\in\Sigma_y$ to $q\in\Sigma_{y+\dy}$ displayed in figure \ref{fig:ADM}.

The Lie derivatives along $D_y$ may now be defined from first principles using the Wilson lines as:
\begin{equation}\label{y deriv}
    \mathcal{L}_{D_y} T_{y} = \lim_{\dy\rightarrow0} \frac{W(y,y+\dy)T_{y+\dy} - T_y}{\dy} \; ,
\end{equation}
which we then discretise as finite difference expressions between neighbouring sites:
\begin{equation}
    \mathcal{L}_{D_y} T \rightarrow \frac{W_{(i, i+1)} T_{i+1} - T_i}{\dy} \; .
\end{equation}
The discretised Wilson line $W_{(i,i+1)}$ is hence an operator that lives on the link between sites $i$ and $i+1$ in our theory graph, and its operation is to pull back tensors from site $i$ to site $i+1$:
\begin{equation}
    W_{(i,i+1)} T_{i+1} = T_i \circ Y_{(i, i+1)} \; .
\end{equation}
The collection of fields $Y^\mu_{(i,i+1)}$, which are the discrete versions of the $Y^\mu(y',y)$ defined in Eq. \eqref{continuum stuckelberg}, are precisely our Stückelberg fields of section \ref{Sec:stuckelberg}! Comparing Eq. \eqref{Goldstones} for the Stückelberg fields with our continuum expression $Y^\mu(y,y+\dy) = x^\mu + N^\mu\dy$, one sees that the shift vectors $N_i^\mu$ on each site\footnote{But wait, didn't these Goldstone bosons live on the links in the deconstructed theory, rather than on particular sites? Well, yes -- we will have more to say about this in section \ref{Sec:new deconstruction}; the early spoiler is that the Goldstone bosons $\pi^\mu_{(i,i+1)}$ are actually better represented by the \emph{average} of the shifts on neighbouring sites, $(N^\mu_i+N_{i+1}^\mu)/2$.} become the Goldstone bosons of the broken diffeomorphisms in the deconstructed theory, and that choosing unitary gauge for the Stückelberg fields is the same as gauge fixing $N^\mu=0$ in the continuum theory. Indeed, the ability to choose $N^\mu=0$ in 5-dimensional GR without changing the physics is reflected in the deconstructed theory through the fact that the Stückelberg equations of motion are already encoded in the unitary gauge Bianchi constraint (c.f. Eq. \eqref{stuckelberg bianchi}).

Still, there is some freedom in how one chooses to apply Eq. \eqref{y deriv} to discretise the 5-dimensional action \eqref{5D GR} -- namely, does one choose to discretise the metrics or the vielbeins? Pre-dRGT, it was the metrics themselves that were discretised \cite{discrete_grav_dims,Schwartz_construction,multigrav_from_ED,Deffayet_deconstruction_review,solutions_in_multigrav_deffayet}, but the interactions in the resulting multi-metric theories were not of the dRGT type, and the theories consequently possessed ghosts. 

After the development of ghost free multi-gravity, it was shown in \cite{deconstructing_dims} that, by discretising the 5-dimensional vielbeins instead of the metrics, this problem is fixed, as one generates ghost free interactions upon deconstruction that are precisely of the form \eqref{MultigravPotential} -- \emph{provided that one gauge fixes the lapse to $\mathcal{N}=1$ everywhere before discretising}. In terms of the vielbeins, the extrinsic curvature reads \cite{deconstructing_dims}:
\begin{equation}\label{ext curvature vielbein}
    K_{\mu\nu} = \frac{1}{2\mathcal{N}}\eta_{ab}\left(e_{\mu}^{\;a} \mathcal{L}_{D_y}e_{\nu}^{\;b} + e_{\nu}^{\;b} \mathcal{L}_{D_y}e_{\mu}^{\;a} \right) \; ,
\end{equation}
so only when one has $\mathcal{N}=1$ do we obtain upon discretisation that $K_{\mu\nu}$ should be replaced by:
\begin{equation}\label{K to vielbein}
    K_{\mu\nu} \rightarrow \frac12 \left(e^{(i)}_{\mu a}\frac{\tilde{e}^{(i)a}_{\nu} - \vbein{i}{\nu}{a}}{\dy} - e^{(i)}_{\nu a}\frac{\tilde{e}^{(i)a}_{\mu} - \vbein{i}{\mu}{a}}{\dy}\right) \; .
\end{equation}
Powers of this expression -- owing to the Deser-van Nieuwenhuisen condition \eqref{SymmetricVierbeinCondition} -- are exactly the sorts of terms that appear upon expanding out the multi-vielbein potential \eqref{MultigravPotential}, including the Stückelberg fields (the $\tilde{e}_\mu^{(i)a}$ are as defined in Eq. \eqref{vielbein Stuckelberg}). 

This fixing of the lapse turns out to be the harbinger of much strife regarding the behaviour of the deconstructed theory and the validity of its continuum limit, upon which we shall elaborate in section \ref{sec:fixing lapse}. However, already one can see that it is a problem, since the structure the lapse provides in the extra dimensional theory, namely diffeomorphism invariance in the $y$-direction, as well as the 5-dimensional Hamiltonian constraint, will be missing in the resulting 4-dimensional multi-gravity theory. Thus, GR \emph{cannot} be a completely consistent continuum limit of standard multi-gravity, as the continuum theory must break diffeomorphisms in the $y$-direction and hence contain an additional degree of freedom \cite{deconstructing_dims}. Conversely, standard multi-gravity \emph{cannot} consistently arise from the deconstruction of higher dimensional GR; the resulting 4-dimensional theory should be something else, although it should still be closely related to standard multi-gravity.

\subsubsection{`Bottom-up' approach}\label{sec:bottom up}

The loss of full 5-dimensional diffeomorphism invariance in the continuum theory motivates a different approach to deconstruction, where one starts with standard multi-gravity in 4-dimensions and attempts to build upwards towards whatever 5-dimensional gravitational theory it corresponds to in the continuum (which cannot be GR, by the argument above). This `bottom-up' approach was developed in \cite{Deconstructing_CW,ClockworkCosmo}, although it was not spoken of in these terms, as the authors (myself included) were focused more on an application of multi-gravity to the Higgs hierarchy problem (see also \cite{ClockworkGrav}).

The idea is to explicitly expand out the interaction potential \eqref{MultigravPotential} in vielbein form and identify the various contractions of the extrinsic curvature that appear, given that they can be expressed in terms of the vielbeins using Eq. \eqref{K to vielbein}. The original papers performed this calculation in unitary gauge for the Stückelberg fields, so all instances of $\tilde{e}^{(i)a}_\mu$ in both Eq. \eqref{MultigravPotential} and Eq. \eqref{K to vielbein} were replaced by $\vbein{i+1}{\mu}{a}$, but the procedure works equally well if one keeps the Stückelberg fields in, as we will do here. With chain-type interactions, the 4-dimensional multi-vielbein potential expands out as:
\begin{equation}\label{expanded potential}
    \begin{split}
        I_V = -24 \sum_{i} \int \dd[4]x \sqrt{\abs{e_{(i)}}}\,  \big( T_{iiii} &+ T_{iii,i+1} \tilde{e}_\mu^{(i)a} \ivbein{i}{\mu}{a}
        \\
        &+ T_{ii,i+1,i+1} \tilde{e}_{[\mu}^{(i)a}\tilde{e}_{\nu]}^{(i)b} \ivbein{i}{\mu}{a}\ivbein{i}{\nu}{b}
        \\
        &+ T_{i,i+1,i+1,i+1} \tilde{e}_{[\mu}^{(i)a}\tilde{e}_{\nu}^{(i)b}\tilde{e}_{\rho]}^{(i)c} \ivbein{i}{\mu}{a}\ivbein{i}{\nu}{b}\ivbein{i}{\rho}{c} \big) \; ,
    \end{split}
\end{equation}
where the square brackets around some of the indices denote their antisymmetrisation i.e. $A_{[\mu}B_{\nu]} = \frac{1}{2!}(A_\mu B_\nu - A_\nu B_\mu)$. Using the Deser-van Niewenhuisen condition \eqref{SymmetricVierbeinCondition}, together with Eq. \eqref{K to vielbein} and the symmetry of the $T_{ijkl}$ coefficients, after a bit of work one may express this potential in the following way:
\begin{equation}\label{potential in K}
    \begin{split}
        I_V = -24 \sum_{i} \int \dd[4]x \dy \sqrt{-\det g_{(i)}}\,  &\bigg[ \frac{1}{\dy} \left(\Tz + 4\Ti +6\Tii +4\Tiii\right)
        \\
        &+ \left(\Ti + 3\Tii - 3\Tiii\right) K
        \\
        &+ \dy \left(\Tii + 2\Tiii \right) K_{(2)}
        \\
        &+ \dy^2 \Tiii K_{(3)} \bigg] \; ,
    \end{split}
\end{equation}
where we have defined the following scalars from the extrinsic curvature:
\begin{align}
    K_{(2)} &= \delta^\mu_{[\alpha}\delta^\nu_{\beta]} K^\alpha_{\;\mu} K^\beta_{\;\nu} \; ,
    \\
    K_{(3)} &= \delta^\mu_{[\alpha}\delta^\nu_{\beta}\delta^\rho_{\gamma]} K^\alpha_{\;\mu} K^\beta_{\;\nu} K^\gamma_{\;\rho} \; .
\end{align}
Next, we note that the kinetic term simply becomes:
\begin{equation}\label{wrong kinetic}
    I_K = \sum_i \int \dd[4]x\dy\sqrt{-\det g_{(i)}} \left[\frac{M_{(4)}^2}{2\dy} \Rsi{i} - \frac{2\Lambda_4}{\dy}\right] \; ,
\end{equation}
where we have assumed for simplicity that all of the bare gravitational couplings are the same, $M_i=M_{(4)}$. We have also explicitly included a bare cosmological constant term, $\Lambda_4$, that we have separated out from the rest of $\Tz$ for convenience in writing down the continuum action, which now follows immediately by replacing $\sum_i\dy$ with an integral over $\dd y$ and including the boundary term \cite{Deconstructing_CW,ClockworkCosmo}:
\begin{equation}\label{wrong continuum}
\begin{split}
    I_{(5)} &= \int \dd[4]x\dd y \sqrt{-g}\left[\frac{M_{(5)}^3}{2}R_{(4)} - 2\Lambda_5 + \alpha_1 M_{(5)}^4 K + (\alpha_2+1) M_{(5)}^3 K_{(2)} + \alpha_3 M_{(5)}^2 K_{(3)}\right]
    \\
    &+ \int_{\partial\mathcal{M}} M_{(5)}^3 K \; ,
\end{split}
\end{equation}
where one has the identifications $M_{(5)}^3 = M_{(4)}^2/\dy$ and $\Lambda_5 = \Lambda_4/\dy$, and the $T_{ijkl}$ are related to the $\alpha_{1,2,3}$ by\footnote{Note that this means the first term in Eq. \eqref{potential in K} with the prefactor $1/\dy$ vanishes; if this term did not vanish, it would contribute a cosmological constant to the continuum theory, but we have already separated this contribution out as $\Lambda_{4,5}$ for convenience.}:
\begin{align}
    24\Tz &= 28\alpha_3 \frac{M^2_{(5)}}{\delta y^2} - 6 (\alpha_2+1) \frac{M^3_{(5)}}{\delta y} + 4\alpha_1 M^4_{(5)}\label{Tz} \; ,
    \\
    24\Ti &= -9\alpha_3 \frac{M^2_{(5)}}{\delta y^2} +3 (\alpha_2+1) \frac{M^3_{(5)}}{\delta y} - \alpha_1 M^4_{(5)}\label{Ti} \; ,
    \\
    24\Tii &= 2\alpha_3 \frac{M^2_{(5)}}{\delta y^2} - (\alpha_2+1) \frac{M^3_{(5)}}{\delta y}\label{Tii} \; ,
    \\
    24\Tiii &= -\alpha_3 \frac{M^2_{(5)}}{\delta y^2}\label{Tiii} \; .
\end{align}

If one tunes the $T_{ijkl}$ such that $\alpha_1=\alpha_2=\alpha_3=0$, the continuum action \eqref{wrong continuum} reads exactly the same as the 5-dimensional GR action \eqref{5D GR} in 4+1 ADM variables, written in a gauge where the lapse is $\mathcal{N}=1$. Naively, in \cite{Deconstructing_CW,ClockworkCosmo} we used this to repackage the $R_{(4)}$ and $K_{(2)}$ terms into the 5-dimensional Ricci scalar in this gauge, as well as to write $\sqrt{-g}=\sqrt{-G}$, also true for $\mathcal{N}=1$ and $N^\mu=0$. However, one should not do this, \emph{as the two theories are not the same}; the lapse \emph{never enters} the continuum theory defined by \eqref{wrong continuum}, and although it appears so, it has \emph{not} been simply gauged away. This is a subtle point, but it is an important one, as it again implies that GR could never have been a sensible continuum limit of standard multi-gravity, and motivates us to develop a better deconstruction procedure. 

Nevertheless, we were on the right track with the bottom-up approach, as we will see in section \ref{Sec:new deconstruction} that it helps to inform what the correct 4-dimensional theory should be if one wishes to recover full GR (with the lapse) in 5-dimensions upon taking the continuum limit. Before that, we would like to discuss in more detail what exactly the problems are with the missing lapse function, following the discussions first outlined in \cite{deconstructing_dims}.

\subsection{Why fixing the lapse is bad}\label{sec:fixing lapse}

Already we have seen some of the issues with this in the latter two subsections: the missing lapse means that the continuum theory has no diffeomorphism invariance in the $y$-direction, nor does it possess a 5-dimensional Hamiltonian constraint. This is obviously bad, and presumably renders the continuum limit of standard multi-gravity ill-defined. However, one can see that issues related to the missing lapse arise already at the level of the deconstructed theory; they manifest in the strange strong coupling properties of the known ghost free theories of massive spin-2 fields \cite{EFTforMGs,deconstructing_dims,SC_and_graph_structure,complete_decoupling_limit,decoupling_limit_multigravity}.

To determine where multi-gravity becomes strongly coupled, one makes use of the Goldstone boson equivalence theorem to relate the dynamics of the Stückelberg fields to the different helicity states of the massive spin-2 fields, upon taking the so-called `decoupling limit'. Precisely, assuming that there exists a background wherein all the metrics are Minkowski (which always exists providing that one chooses the interaction coefficients to make it so \cite{consistent_spin2,prop_bg_multigrav,BH2}) and expanding the action around it, one generates interactions between the different metric perturbations $h^{(i)}_{\mu\nu}$ and Stückelberg perturbations $\pi^\mu_{(j,i)}$, which we recall can be decomposed into $\pi^\mu_{(j,i)}= A^\mu_{(j,i)}+\partial^\mu\phi_{(j,i)}$. The least suppressed interactions that are generated come with some scale usually denoted $\Lambda_3$, which depends on the spin-2 masses $m_i$ (implicitly related to the interaction coefficients) and on $M_{(4)}$; the decoupling limit is taken by simultaneously sending $m_i\rightarrow0$, $M_{(4)}\rightarrow\infty$ while keeping $\Lambda_3$ fixed, which isolates these interactions and ensures that the fields $h_{\mu\nu}$, $A_\mu$ and $\phi$ really do represent the helicity-2, helicity-1 and helicity-0 modes of the massive gravitons in this limit \cite{SC_and_graph_structure,decoupling_limit_multigravity}.

After diagonalising all kinetic mixings in the decoupling limit, for the case of chain interactions, one finds that the interaction appearing at the lowest scale is schematically $h(\partial\phi)^2$, which is suppressed by the scale \cite{SC_and_graph_structure,deconstructing_dims}:
\begin{equation}
    \Lambda_{\text{SC}} = (m_1^2 \Mp)^{\frac13} \; ,
\end{equation}
where $\Mp=\sqrt{N}M_{(4)}$ is the effective Planck scale of the massless mode around the Minkowski vacuum \cite{BHs_multigrav} and $m_1$ is the mass of lightest massive mode. This is the energy scale at which the theory becomes strongly coupled, and it can be much lower than the Planck scale depending on the masses of the gravitons\footnote{Graviton masses in multi-gravity theories are observationally constrained to be either very heavy ($\sim$TeV and above) or ultra-light (sub $\sim$$10^{-32}$eV) \cite{Solar_sys_tests,Graviton_mass_bounds,heavy_spin2_DM}. Heavy gravitons are strongly coupled above energies we can currently reach with experiments, so this is fine in an EFT sense; light gravitons, however, become strongly coupled at extremely low energies, and in fact the smallness of the scale $\Lambda_{\text{SC}}$ for ultra-light gravitons is crucial in establishing the onset of the Vainshtein mechanism that screens them from our view \cite{Vainshtein,Recovering_GR,Vainshtein_decoupling_limit,Vainshtein_recovery,Vainshtein_review}.}.

If one chooses to pick their interaction coefficients in such a way that would naively recover GR in 5-dimensions (of course, we know now that this is impossible to achieve in reality) i.e.
\begin{align}
    24\Tz &= -\frac{6M_{(5)}^3}{\dy} \; ,
    \\
    24\Ti &= \frac{3M_{(5)}^3}{\dy} \; ,
    \\
    24\Tii &= -\frac{M_{(5)}^3}{\dy} \; ,
    \\
    \Tiii &= 0 \; ,
\end{align}
then one can show explicitly \cite{BHs_multigrav} that the graviton masses upon taking the continuum limit are just those of the standard KK graviton spectrum, $m_n = n\pi/L$, while the effective Planck scale obeys its usual relationship with the size of the extra dimension, ${M_{\text{Pl}}^2=NM_{(4)}^2=NM_{(5)}^3\delta y=M^3_{(5)}L}$. Thus, for this particular model, the theory becomes strongly coupled at the scale:
\begin{equation}
    \Lambda_{\text{SC}} \sim \left(\frac{M_{(5)}}{L}\right)^{\frac12} \; ,
\end{equation}
which exhibits a bizarre UV-IR mixing through its dependence on $L$, the size of the extra dimension, and is \emph{lower} than the 5-dimensional Planck scale $M_{(5)}$. If the multi-gravity theory really descended from 5-dimensional GR, one would naively expect it to be valid all the way up to $M_{(5)}$, which is yet another way of seeing that standard multi-gravity cannot arise from deconstructing GR.

Moreover, the appearance of this low, IR-dependent strong-coupling scale in multi-gravity can be directly related to the fixing of the lapse in the continuum theory \cite{deconstructing_dims}. The dangerous interactions in the deconstructed theory, as we saw, involve terms coupling the helicity-2 ($h_{\mu\nu}$) and helicity-0 ($\phi$) graviton modes, which arise from the decomposition of the Stückelberg perturbations. In the continuum, we saw that the analogue of the Stückelberg perturbations is the shift vector, which we can similarly decompose as $N^\mu=A^\mu+\partial^\mu\phi$ in the background $g_{\mu\nu}=\eta_{\mu\nu}$. The authors of \cite{deconstructing_dims} showed that, in the gauge where the lapse is fixed to $\mathcal{N}=1$, exactly the same $h(\partial\phi)^2$ interactions appear in the 5-dimensional theory, but in this gauge the scalar $\phi$ has a kinetic mixing with $\partial_y g_{\mu\nu}$ so has to be ``canonically normalised'' by the formal replacement $\phi\rightarrow(1/\partial_y)\phi$. Since $\partial_y$ can be made arbitrarily small (at least, up to $\sim1/M_{(5)}$), the scale at which the $h(\partial\phi)^2$ interactions become strongly coupled too can be decreased at will, and we apparently have the same issue as in the deconstructed theory. In 5-dimensional GR, we know that this is just a gauge artifact that can be resolved by reinserting the lapse; however, upon deconstruction we do not have this luxury anymore as the lapse was never there: what was a gauge choice in 5-dimensions has been converted into a genuine physical scale in 4-dimensions! Thus, it was conjectured that whatever the 4-dimensional theory is that becomes 5-dimensional GR in its continuum limit, it should not exhibit such a low strong coupling scale. 

We are not going to determine the strong coupling scale of said theory in this work, saving it instead for a future paper, as we believe the calculation is important enough to warrant its own article. Nevertheless, we note that the determination of $\Lambda_{\text{SC}}$ in our new theory is a necessary step to confirm that the continuum limit is well-defined up to the right energy scale, as we develop our improved deconstruction procedure.

\section{A more complete deconstruction procedure}\label{Sec:new deconstruction}

The goal now is to modify the deconstruction procedure discussed in section \ref{Sec:old deconstruction} in such a way that the lapse remains free, in order to circumvent the problems that ensue when one fixes $\mathcal{N}=1$ before discretising the extra dimension. Thankfully, having already spent a great deal of time developing the necessary technology to achieve this, the path forwards should be reasonably clear. 

Firstly, it is immediately obvious that the fields contained in the deconstructed theory should comprise: a collection of vielbeins/metrics corresponding to the induced hypersurface metrics, owing to Eq. \eqref{discrete metrics}; a collection of Stückelberg fields related to the discretised shift vectors, owing to Eq. \eqref{discrete shifts}; and finally a collection of scalars corresponding to the discretised lapse, owing to Eq. \eqref{discrete lapses}. 

The generic structure of the interactions between the different vielbeins should be the same as in standard multi-gravity, since the manner in which the vielbeins enter the discretised expression for the extrinsic curvature is unchanged by the presence of the lapse (c.f. Eq. \eqref{ext curvature vielbein}). Therefore, the only way in which the new scalars are permitted to enter the potential of the deconstructed multi-gravity theory is implicitly through the coefficients $T_{ijkl}$, otherwise the wedge product form of the multi-vielbein potential \eqref{MultigravPotential} would be disrupted. Hence, one should promote these coefficients to functions of the scalars\footnote{In fact, to be completely general, we could in principle allow the $\tilde{T}_{ijkl}$ to be a function of both the scalars $\mathcal{N}_i$ and their 4-dimensional derivatives, $\partial\mathcal{N}_i$, which would correspond to giving the scalars kinetic terms in the deconstructed theory. However, only $\tilde{T}_{iiii}$ would be permitted to depend on the derivatives in this manner, as terms like $\nabla^{(i)}\mathcal{N}_i\nabla^{(j)}\mathcal{N}_i$ arising from more general $\tilde{T}_{ijkl}$ would violate the no-go theorems that forbid matter coupling to multiple vielbeins simultaneously \cite{ghost_doubly_coupled,on_matter_couplings,ghosts_matter_couplings_rev}. Therefore, we will leave all these derivative couplings out for now, but in section \ref{Sec:MMST} we will put the safe ones back in.}, $T_{ijkl}\rightarrow\tilde{T}_{ijkl}(\bm{\mathcal{N}})$, where $\bm{\mathcal{N}}=\{\mathcal{N}_i,\mathcal{N}_j,\mathcal{N}_k,\mathcal{N}_l\}$, which reduce back to the standard $T_{ijkl}$ whenever all $\mathcal{N}_i=1$. That the interaction structure is unchanged also informs us that the Stückelberg mechanism should operate in the same way as in standard multi-gravity, so one may choose to work in unitary gauge without issue if one so wishes. 

These realisations motivate us to look for a deconstructed theory whose continuum limit is of a similar form to Eq. \eqref{wrong continuum}, but tweaked slightly to reintroduce the arbitrary lapse function; namely, we should look for a continuum action of the form:
\begin{equation}\label{right ADM continuum}
\begin{split}
    I_{(5)} = \int \dd[4]x\dd y \sqrt{-g}\mathcal{N}&\Bigg[\frac{M_{(5)}^3}{2}R_{(4)} - 2\Lambda_5 + \alpha_1(\mathcal{N}) M_{(5)}^4 K \\
    &+ \left(\alpha_2(\mathcal{N})+1\right) M_{(5)}^3 K_{(2)} + \alpha_3(\mathcal{N}) M_{(5)}^2 K_{(3)}\Bigg]
    + \int_{\partial\mathcal{M}} M_{(5)}^3 K \; .
\end{split}
\end{equation}
In general, since the 4-dimensional interaction coefficients $\tilde{T}_{ijkl}$ are now functions of the scalars $\mathcal{N}_i$, the $\alpha_{1,2,3}$ coefficients constructed from them may also depend on the local value of the lapse in the extra dimension. 

This theory still possesses some strange aspects when $\alpha_{1,2,3}\neq 0$: the presence of the additional extrinsic curvature contributions again explicitly breaks diffeomorphism invariance in the $y$-direction, and it is unclear whether these extra terms can in general be reinterpreted in terms of 5-dimensional curvature quantities (or as a new scalar degree of freedom as in \cite{Deconstructing_CW}) for arbitrary $\alpha_{1,2,3}(\mathcal{N})$. Nevertheless, the return of the lapse allows one to now repackage $R_{(4)}$ and $K_{(2)}$ into $R_{(5)}$, as well as $\sqrt{-g}\mathcal{N}$ into $\sqrt{-G}$, so that the 5-dimensional action reads:
\begin{equation}\label{right continuum}
    \begin{split}
    I_{(5)} = \int \dd[5]x \sqrt{-G}&\Bigg[\frac{M_{(5)}^3}{2}R_{(5)} - 2\Lambda_5 + \alpha_1(\mathcal{N}) M_{(5)}^4 K \\
    &+ \alpha_2(\mathcal{N}) M_{(5)}^3 K_{(2)} + \alpha_3(\mathcal{N}) M_{(5)}^2 K_{(3)}\Bigg] + \int_{\partial\mathcal{M}} M_{(5)}^3 K \; .
\end{split}
\end{equation}
Hence, when the interaction coefficients are tuned to give $\alpha_1=\alpha_2=\alpha_3=0$, this continuum theory really does become 5-dimensional GR, unlike in the scenario from the previous section where the lapse was missing. This suggests we may be getting somewhere -- we just need to determine the nature of the deconstructed 4-dimensional theory that has Eq. \eqref{right continuum} as its continuum limit.

To get there, we shall use the intuition we have developed from the bottom-up approach to deconstruction. To obtain the kinetic part is easy: one should take in the deconstructed theory the following kinetic term:
\begin{equation}
    I_K = \sum_{i=0}^{N-1}\int \dd[4]x \sqrt{-\det g_{(i)}} \,\mathcal{N}_i \left[\frac{M_{(4)}^2}{2} R_{(i)} - 2\Lambda_4\right] \; ,
\end{equation}
which one may write, in analogy with Eq. \eqref{wrong kinetic}, as:
\begin{equation}
    I_K = \sum_{i=0}^{N-1} \int \dd[4]x\dy\sqrt{-\det g_{(i)}} \,\mathcal{N}_i \left[\frac{M_{(4)}^2}{2\dy} R_{(i)} - \frac{2\Lambda_4}{\dy}\right] \; .
\end{equation}
This trivially recovers the first two terms of Eq. \eqref{right ADM continuum} in the continuum limit, with the identifications $M_{(5)}^3 = M_{(4)}^2/\dy$ and $\Lambda_5 = \Lambda_4/\dy$.

To obtain the right potential requires more thought. Previously, we expanded the standard multi-vielbein potential \eqref{MultigravPotential} and identified terms of the form \eqref{K to vielbein} with the extrinsic curvature of hypersurfaces in the extra dimension. However, Eq. \eqref{K to vielbein} is written in a gauge with $\mathcal{N}=1$, so we now want to replace it with a discretised version of Eq. \eqref{ext curvature vielbein}, where the lapse remains arbitrary. One must take care here to ensure that the choice of discretisation procedure does not single out any distinguished site(s) in the deconstructed theory, since the equation for the lapse in the higher dimensional theory holds throughout all of spacetime. In practice, this means that one should discretise the $1/2\mathcal{N}$ factor in Eq. \eqref{ext curvature vielbein} using some combination of the scalars $\mathcal{N}_i$ and $\mathcal{N}_{i+1}$ that makes the whole expession one that lives firmly on the link between sites $i$ and $i+1$, rather than being weighted to either side. A natural and minimal choice is simply the average of these scalars, $\langle\mathcal{N}_i\rangle=(\mathcal{N}_i+\mathcal{N}_{i+1})/2$, which dictates that one should discretise the extrinsic curvature as:
\begin{equation}\label{better K discretisation}
    K_{\mu\nu}\rightarrow \frac{1}{\mathcal{N}_i+\mathcal{N}_{i+1}}\left(e^{(i)}_{\mu a}\frac{\tilde{e}^{(i)a}_{\nu} - \vbein{i}{\nu}{a}}{\dy} - e^{(i)}_{\nu a}\frac{\tilde{e}^{(i)a}_{\mu} - \vbein{i}{\mu}{a}}{\dy}\right) \; ,
\end{equation}
for which the updated version of Eq. \eqref{potential in K} becomes:
\begin{equation}\label{naive continuum}
    \begin{split}
        I_V = -24 \sum_{i} \int \dd[4]x \dy \sqrt{-\det g_{(i)}}\, \langle\mathcal{N}_i\rangle &\bigg[ \left(\tilde{T}_{iii,i+1} + 3\tilde{T}_{ii,i+1,i+1} - 3\Tiii\right) K
        \\
        &+ \langle\mathcal{N}_i\rangle\dy \left(\tilde{T}_{ii,i+1,i+1} + 2\tilde{T}_{i,i+1,i+1,i+1} \right) K_{(2)}
        \\
        &+ \langle\mathcal{N}_i\rangle^2\dy^2 \tilde{T}_{i,i+1,i+1,i+1} K_{(3)} \bigg] \; ,
    \end{split}
\end{equation}
after accounting for the vanishing cosmological constant term that has already been extracted as $\Lambda_4$ (c.f. Eq. \eqref{potential in K}). In the continuum limit, this indeed recovers the remaining few terms in the action \eqref{right continuum}, where the potential coefficients are expressed in terms of $\alpha_{1,2,3}(\bm{\mathcal{N}})$ as:
\begin{align}
    24\tilde{T}_{iiii} &= 28\alpha_3(\bm{\mathcal{N}}) \frac{M^2_{(5)}}{\langle\mathcal{N}_i\rangle^2\delta y^2} - 6 (\alpha_2(\bm{\mathcal{N}})+1) \frac{M^3_{(5)}}{\langle\mathcal{N}_i\rangle\delta y} + 4\alpha_1(\bm{\mathcal{N}}) M^4_{(5)} \; ,\label{Tz tilde}
    \\
    24\tilde{T}_{iii,i+1} &= -9\alpha_3(\bm{\mathcal{N}}) \frac{M^2_{(5)}}{\langle\mathcal{N}_i\rangle^2\delta y^2} +3 (\alpha_2(\bm{\mathcal{N}})+1) \frac{M^3_{(5)}}{\langle\mathcal{N}_i\rangle\delta y} - \alpha_1(\bm{\mathcal{N}}) M^4_{(5)} \; ,
    \\
    24\tilde{T}_{ii,i+1,i+1} &= 2\alpha_3(\bm{\mathcal{N}}) \frac{M^2_{(5)}}{\langle\mathcal{N}_i\rangle^2\delta y^2} - (\alpha_2(\bm{\mathcal{N}})+1) \frac{M^3_{(5)}}{\langle\mathcal{N}_i\rangle\delta y} \; ,
    \\
    24\tilde{T}_{i,i+1,i+1,i+1} &= -\alpha_3(\bm{\mathcal{N}}) \frac{M^2_{(5)}}{\langle\mathcal{N}_i\rangle^2\delta y^2} \; ,\label{Tiii tilde}
\end{align}
generalising Eqs. \eqref{Tz}--\eqref{Tiii}. We note that the $1/\mathcal{N}$ dependence common to all of the interaction coefficients in the GR case where $\alpha_1=\alpha_2=\alpha_3=0$ was also found in the original top-down deconstruction papers by Deffayet and Mourad \cite{multigrav_from_ED,Deffayet_deconstruction_review,solutions_in_multigrav_deffayet}. However, as we mentioned in section \ref{sec:top down}, the multi-metric interactions in those works proved ultimately to be ghostly; our results above are the analogue of theirs for ghost free interactions.

We stress that the means of discretising the extra dimension provided by Eq. \eqref{better K discretisation} is not necessarily the most general procedure one could opt for; the discretisation method is ultimately a choice one makes -- for example, one could use centralised derivatives involving more sites, rather than the simple nearest neighbour derivatives we use here\footnote{In fact, in the original deconstruction papers it was argued that, to account entirely for the structure of the extra dimension, in discretising the $y$-derivatives one should really use a prescription containing \emph{all} of the sites, just having the contribution of the distant sites fall away quickly \cite{Schwartz_construction}. This would lead to a deconstructed theory whose theory graph links every site with every other, which, as we argued, would be ghostly in 4-dimensions. Therefore, we stick with the simple choice of discretisation given by Eq. \eqref{better K discretisation}, which ensures the only interactions in the deconstructed theory are nearest neighbour; we will show that it is still good enough to capture the relevant dynamics in the continuum limit.}. Equally, one could make a different choice than the average of the scalars for how to put the extrinsic curvature on the links between adjacent sites. However, we will soon show that the deconstructed theory we have arrived at by using Eq. \eqref{better K discretisation} is able to recover all of the dynamics and constraints of 5-dimensional brane cosmology upon taking the continuum limit, so the procedure we have introduced seems to be at least a good one.

\subsection{A new kind of theory: multi-gravity with scalar fields}

By developing an improved dimensional deconstruction procedure for 5-dimensional gravity that keeps the lapse free, we have arrived at a new 4-dimensional theory whose action is of the following general form:
\begin{align}
    I &= I_K + I_V + I_M\label{NewAction}
    \\
    I_K &= \sum_{i=0}^{N-1} \int_{\mathcal{M}_4} \mathcal{N}_i\left[ \frac{M_{(4)}^{2}}{2} R^{(i)}_{ab} \wedge \hodge^{(i)} e^{(i)ab}-2\Lambda_4\hodge^{(i)}1\right]
    \\
    I_V &= -\sum_{i,j,k,l=0}^{N-1} \int_{\mathcal{M}_4} \varepsilon_{abcd} {T}_{ijkl}\left(\bm{\mathcal{N}}\right) e^{(i)a}\wedge e^{(j)b}\wedge e^{(k)c} \wedge e^{(l)d} \; ,
\end{align}
where we have dropped the tildes on the $T_{ijkl}$ coefficients for convenience, with the implicit understanding that they now are functions of the scalar fields $\mathcal{N}_i$. The choice for these coefficients that recovers pure 5-dimensional GR in the continuum limit is given by Eqs. \eqref{Tz tilde}--\eqref{Tiii tilde} with $\alpha_1=\alpha_2=\alpha_3=0$.

This is a hybrid of multi-gravity with a Brans-Dicke type scalar-tensor theory, with characteristic parameter $\omega=0$ \cite{BransDicke}. Consequently, for want of a better name, we shall henceforth refer to the theory described by the action \eqref{NewAction} as ``Scalar-Tensor Multi-Gravity'' (or \emph{STMG} for short). In section \ref{Sec:MMST}, we are going to generalise the above action to both arbitrary dimension and more general scalar couplings in the kinetic sector; for now, we will stick with the $\omega=0$ Brans-Dicke subclass, given that we would like to claim it can arise from the deconstruction of 5-dimensional GR.

We note that dRGT-like theories of massive gravity with a single dynamical metric (i.e. $N=1$), but where the graviton mass is dependent on the value of one or more varying scalar fields, have actually already been studied under the name of `mass-varying (or generalised) massive gravity' -- see e.g. references \cite{MVMG,Generalised_massive_grav,Cosmo_generalised_MG}. Likewise, a simple bi-metric, bi-scalar (i.e. $N=2$) example from within the $\omega=0$ Brans-Dicke subclass of STMG theories too has been studied recently in the context of cosmic inflation \cite{Starobinsky_bigravity}. There, the authors constructed a bigravity analogue of the Starobinsky model of inflation \cite{Starobinsky} by extending the kinetic sectors of both metrics by $R_{(i)}^2$ curvature corrections. Upon making a field redefinition, one may reinterpret the additional $R_{(i)}^2$ terms as a pair of scalar fields; the bi-metric Starobinsky action then takes precisely the form of the metric formulation of Eq. \eqref{NewAction} (which we will introduce shortly), with some additional scalar potentials that can be absorbed into $T_{iiii}$. In both mass-varying massive gravity and its bi-metric extension, the theory \emph{remains ghost free}, since the fundamental structure of the multi-vielbein interactions required to exorcise the ghostly degrees of freedom is unchanged from standard multi-gravity; indeed, one can show explicitly via Hamiltonian constraint analysis that the additional scalar fields contribute in a perfectly benign manner to the constraints that kill the ghosts \cite{dR_review,MVMG,Generalised_massive_grav}. For example, the bi-metric Starobinsky model propagates only a massless spin-2 field, a massive spin-2 field, and 2 scalars (for a total of 2+5+1+1=9 degrees of freedom) \cite{Ghost_freedom_quadratic_bigravity}. Importantly, these proofs of ghost freedom do not depend on the specific form that the $T_{ijkl}(\bm{\mathcal{N}})$ take as a function of the scalars, so we expect that the generic STMG theory defined by Eq. \eqref{NewAction}, with an arbitrary number of fields, should also remain ghost free, provided that the non-vanishing $T_{ijkl}$ obey the permitted ghost free structures that we outlined in section \ref{Sec:vielbein}. Precisely, the propagating degrees of freedom within the theory should generically comprise: 1 massless spin-2 field, $N-1$ massive spin-2 fields, and $N$ scalars, with no ghosts.  As was the case in standard multi-gravity, one may track the helicity states of the massive spin-2 fields explicitly by replacing all instances of $e^{(j)a}_\mu$ in the potential with $\tilde{e}^{(i)a}_\mu$, given in terms of the Stückelberg fields as in Eq. \eqref{vielbein Stuckelberg}, thereby restoring invariance to the theory under the product of 4-dimensional diffeomorphisms $\GC{1}\times\hdots\times\GC{N}$\footnote{Note that the scalars transform under these diffeomorphisms as $\delta\mathcal{N}_i = \xi_i^\mu\partial_\mu\mathcal{N}_i$ \cite{multigrav_from_ED,Deffayet_deconstruction_review}.}.

Although we will not analyse the spectrum explicitly in this work, it behoves us to comment on how this degree of freedom count makes sense when one considers the more standard Kaluza-Klein story of integrating out the extra compact dimension -- related to deconstruction by a discrete Fourier transform of the vielbeins \cite{deconstructing_dims} -- in the case where the ${T}_{ijkl}(\bm{\mathcal{N}})$ are chosen to correspond to GR in the continuum limit. In the standard KK picture, it is well-understood (see e.g. \cite{Ortin}) that the 5 degrees of freedom of the 5-dimensional massless graviton should translate into a 4-dimensional spectrum containing: in the massless sector, a single massless spin-2 field with 2 dof, a massless vector (the graviphoton) with 2 dof and a massless scalar (the radion) with 1 dof, for a total of 5 massless dof; and in the massive sector, a finite tower of massive spin-2 fields with masses up to the cutoff of the theory, each propagating 5 dof. 

In the original deconstruction papers by Deffayet and Mourad, it was shown explicitly that in the case where the continuum theory is pure GR in 5-dimensions, an additional discrete symmetry is present in the deconstructed theory, inherited from diffeomorphism invariance in the $y$-direction in the continuum, whose influence is to kill off all of the massive scalar modes, leaving only the massless radion \cite{multigrav_from_ED,Deffayet_deconstruction_review}. Although their deconstructed theory turned out to be ghostly, we expect an analogue of this symmetry to be present in our ghost free deconstruction too, for the particular choice of $T_{ijkl}(\bm{\mathcal{N}})$ that corresponds to pure GR in the continuum. Alternatively, one may choose to keep track of the massless scalar directly from the outset, as was suggested in \cite{deconstructing_dims}, by making the lapse a function of the 4-dimensional coordinates only, $\mathcal{N}(x,y)=\mathcal{N}(x)$, so that all of the $\mathcal{N}_i$ are actually just the same field i.e. the radion. Similarly, the massless vector may be accounted for by switching on a $y$-independent component of the shift vectors, $e_\mu^{\;5}(x)$ (the $y$-dependent parts then become the Stückelberg vectors of the massive spin-2 fields, as we saw) \cite{deconstructing_dims}. 

In the case of arbitrary $T_{ijkl}(\bm{\mathcal{N}})$, the continuum theory is no longer pure GR, as diffeomorphism invariance in the $y$-direction is broken. Therefore, we expect that the continuum theory possesses an additional degree of freedom beyond the 5 of the 5-dimensional massless graviton, as in \cite{Deconstructing_CW}. In the deconstructed theory, this extra degree of freedom cascades down to become our additional scalar fields, which are no longer killed off by any miraculous discrete symmetries. We will develop everything from this point onwards for STMG theory with generic interaction coefficients, so in principle the extra scalars will all be present, but keep in mind that only the massless combination should survive in the case where the continuum limit is pure 5-dimensional GR.

If a particular STMG model has only pairwise interactions, which it must if it is to arise from deconstruction, as we saw during section \ref{Sec:old deconstruction}, then the action \eqref{NewAction} also possesses an equivalent description in the metric formalism that reads:
\begin{align}
    I &= I_K + I_V + I_M\label{NewActionMetric}
    \\
    I_K &= \sum_{i=0}^{N-1} \int \dd[4]x\sqrt{-\det g_{(i)}} \,\mathcal{N}_i\left( \frac{M_{(4)}^{2}}{2}R_{(i)} - 2\Lambda_4 \right)
    \\
    I_V &= -\sum_{i,j} \int \dd[4]x \sqrt{-\det g_{(i)}} \sum_{m=0}^{4} \beta_m^{(i,j)}\left(\bm{\mathcal{N}}\right) e_m(S_{i\rightarrow j})  \; ,
\end{align}
where now the $\beta_m^{(i,j)}(\bm{\mathcal{N}})$ coefficients are assumed to depend on the scalars $\mathcal{N}_i$, but otherwise are still are related to the $T_{ijkl}(\bm{\mathcal{N}})$ by Eqs. \eqref{betas1} and \eqref{betas2}, and reduce to the standard multi-metric $\beta_m^{(i,j)}$ when all $\mathcal{N}_i=1$. As we mentioned in section \ref{Sec:review}, the metric formalism is very useful because it facilitates computation of the field equations, so we will use this formalism going forwards. We will also choose to work throughout this section in unitary gauge for the Stückelberg fields for simplicity, since we know that their field equations are already encoded in the Bianchi constraint. Still, if one wishes to include them explicitly, one should simply make the replacement $S_{i\rightarrow j}\rightarrow \tilde{S}_{i\rightarrow j}$ given by Eq. \eqref{Stuckelberg metric}.

The field equations in unitary gauge follow from the action \eqref{NewActionMetric} by varying with respect to the independent fields $\gi{i}$ and $\mathcal{N}_i$. Varying with respect to the metrics gives the dynamical field equations for $\gi{i}$:
\begin{equation}\label{NewFieldEqs}
\boxed{
    \mathcal{N}_i \left[M_{(4)}^2 G^{(i)}_{\mu\nu} + 2\Lambda_4 \gi{i}\right] + W^{(i)}_{\mu\nu}(\bm{\mathcal{N}}) = T^{(i)}_{\mu\nu} + M_{(4)}^2 \left[\nabla^{(i)}_\mu\nabla^{(i)}_\nu \mathcal{N}_i - \gi{i}\Box^{(i)}\mathcal{N}_i\right] 
} \; ,
\end{equation}
where the $W$-tensor has components:
\begin{equation}\label{NewW}
    \Wi{i}(\bm{\mathcal{N}}) = \sum_j \sum_{m=0}^4(-1)^m\beta_m^{(i,j)}(\bm{\mathcal{N}})Y_{(m)\nu}^\mu(S_{i\rightarrow j})
    +\sum_k\sum_{m=0}^4 (-1)^m\beta_{4-m}^{(k,i)}(\bm{\mathcal{N}})Y_{(m)\nu}^\mu (S_{i\rightarrow k}) \; .
\end{equation}
These reduce to the standard multi-metric equations of motion \eqref{Einstein eqs} when all the $\mathcal{N}_i=1$. In the continuum limit, comparison with the ADM metric \eqref{ADM} says that these equations should become equivalent to the 5-dimensional equations for $g_{\mu\nu}$, the induced metric on the hypersurfaces $\Sigma_y$.

Taking the $i$-th covariant derivative of Eqs. \eqref{NewFieldEqs}, using the Bianchi identity on the Einstein tensors, as well as the fact that \cite{auxiliary_fields_in_gravity}:
\begin{equation}
    \left[\Box^{(i)}\nabla_{\nu}^{(i)} - \nabla_\nu^{(i)}\Box^{(i)}\right]\mathcal{N}_i = R^{(i)\mu}_{\;\;\;\;\;\nu}\nabla^{(i)}_\mu \mathcal{N}_i \; ,
\end{equation}
one finds (assuming the matter coupling is such that all divergences of the energy-momentum tensors vanish individually) that the Bianchi constraint is modified to:
\begin{equation}\label{NewBianchiConstraint}
\boxed{
    \nabla^{(i)}_\mu \Wi{i}(\bm{\mathcal{N}}) = \frac{M_{(4)}^2}{2}R_{(i)}\partial_\nu\mathcal{N}_i
}   \; ,
\end{equation}
which is again equivalent to the standard Bianchi constraint when $\mathcal{N}_i=1$. This should lead to the equation of motion for the shift vector upon taking the continuum limit.

The truly new piece of information we did not have before (i.e. compared against standard multi-gravity) is the variation of the action with respect to the scalars $\mathcal{N}_i$. The field equations one gets by taking these variations are:
\begin{equation}\label{ScalarConstraint}
\boxed{
    \frac{M_{(4)}^2}{2} R_{(i)} - 2\Lambda_4 = \sum_j\sum_{m=0}^4 \frac{\partial\beta_m^{(i,j)}}{\partial\mathcal{N}_i}e_m(S_{i\rightarrow j}) + \sum_k\sum_{m=0}^4 \frac{\partial\beta_{4-m}^{(k,i)}}{\partial\mathcal{N}_i}e_m(S_{i\rightarrow k})
}   \; .
\end{equation}
Although at a first glance these equations appear to be algebraic in the scalar fields, they are in fact not, since the Ricci scalars $R_{(i)}$ carry implicit information regarding the dynamics of each $\mathcal{N}_i$ thanks to the field equations for the metrics. Indeed, by taking the trace of Eq. \eqref{NewFieldEqs} and substituting in for $R_{(i)}$, one may arrive at an equation of the form $\Box^{(i)}\mathcal{N}_i=(\hdots)$ for the scalars, showing that they really are dynamical. We will perform this calculation in section \ref{Sec:MMST} explicitly, but for now Eqs. \eqref{ScalarConstraint} will suffice in the form they are written above, as we will soon show that they are equivalent to the equation for the lapse in the extra dimension, upon taking the continuum limit. 

We stress that even if we are able to find solutions of STMG in which all $\mathcal{N}_i=1$, so that the dynamical equations and Bianchi constraint look exactly the same as they do in standard multi-gravity, the scalar equations must still hold at $\left.\partial\beta_m/\partial\mathcal{N}\right|_{\mathcal{N}=1}$. Standard multi-gravity does not contain any extra scalar fields, so this additional piece of information, present in STMG, is missing in standard multi-gravity. This is the 4-dimensional manifestation of the reason why its continuum limit was unable to recover 5-dimensional GR. The impact of the new scalars is more than just to allow a sensible deconstruction procedure: as we will see in section \ref{Sec:MMST}, the presence of the scalar equations plays a crucial role in determining the kinds of solutions that are permitted in STMG versus in standard multi-gravity. Before we get to that point, however, we would like to make absolutely sure that our new deconstruction procedure is solid, which we will demonstrate by means of a concrete example.

\subsection{Consistency check: recovering brane cosmology}

As a consistency check for our new deconstruction procedure, we are going to demonstrate that the field equations and constraints of STMG really do recover all of the corresponding dynamics and constraints of 5-dimensional GR upon taking the continuum limit. The 5-dimensional model we will choose to show this with is brane cosmology within the Randall-Sundrum-1 (RS1) framework \cite{RS1} (see \cite{Carsten_review,Langlois_review} for reviews).

\subsubsection{Continuum theory}

The RS1 setup takes the 5-dimensional bulk to be described by pure GR, then places 4-dimensional branes at $y=0$ and $y=L$, upon each of which may live a localised energy-momentum tensor, $\Theta_{\mu\nu}$. This energy-momentum tensor is typically separated into a matter part, denoted $\tau_{\mu\nu}$, and a contribution from the brane tension $\sigma$, which acts like a cosmological constant i.e.
\begin{equation}
    \Theta_{\mu\nu}^{(0,L)} = -\sigma_{0,L} \gi{0,L} + \tau^{(0,L)}_{\mu\nu} \; .
\end{equation}
Ordinarily, as we mentioned in section \ref{Sec:old deconstruction}, the extra dimension tends to be orbifolded on $S_1/\mathbb{Z}_2$, so in standard RS1 the branes are two-sided and one can integrate across them. However, since we live instead on the interval $[0,L]$, in our case the branes must be one-sided, end-of-the-world branes located at the boundaries of our 5-dimensional manifold. With this setup, the action for the continuum theory reads:
\begin{equation}
    I_{(5)} = \int_{\mathcal{M}_5}\left(\frac{M_{(5)}^3}{2}R_{(5)} - 2\Lambda_5 \right) + \int_{\partial\mathcal{M}_5} M_{(5)}^3 K + \sum_{i=0,L}\int \dd[4]x \sqrt{-g_{(i)}} \, \mathcal{L}_{m,i} \; .
\end{equation}

In standard GR, variations of the boundary metrics $\gi{0}$ and $\gi{L}$ are typically assumed to vanish. However, here we must allow these variations to be arbitrary, as they correspond to the metrics at either end of the interaction chain in the deconstructed theory, which are dynamical. Hence, there are two sets of field equations: one set for the bulk, owing to the variation of the 5-dimensional Ricci scalar:
\begin{equation}\label{Bulk}
    M_{(5)}^3 G_{MN} = -2\Lambda_5 g_{MN} \; ,
\end{equation}
and one set for the boundary, owing to the variation of the extrinsic curvature:
\begin{align}\label{Boundary1}
    M_{(5)}^3\left.\left(K_{MN}-Kg^{(0)}_{MN}\right)\right|_{y=0} &= - \Theta_{MN}^{(0)}
    \\
    M_{(5)}^3\left.\left(K_{MN}-Kg^{(L)}_{MN}\right)\right|_{y=L} &= + \Theta_{MN}^{(L)} \; ,\label{Boundary2}
\end{align}
where the sign change on the right hand side is due to the change in orientation between the two boundary components.

We note that if one parametrises their 5-dimensional metrics in the 4+1 ADM form of Eq. \eqref{ADM}, then the 55-component of the bulk Einstein equations corresponds to the equation for the lapse function, $\mathcal{N}$, which in pure GR is the Hamiltonian constraint:
\begin{equation}
    M_{(5)}^3G_{55} = -2\Lambda_5 g_{55}\implies  M_{(5)}^3 R_{(4)} - 2\Lambda_5 =  M_{(5)}^3\left(K^2 - K_{\mu\nu}K^{\mu\nu}\right) \; .
\end{equation}
Likewise, the $\mu5$-components of the Einstein equations correspond to the field equations for the shift vector, $N^\mu$, which in pure GR form the momentum constraint:
\begin{equation}
    M_{(5)}^3G_{\mu5} = -2\Lambda_5 g_{\mu5} \implies D^\nu K^\mu_{\;\nu} = D_\nu K \; ,
\end{equation}

One can also understand the boundary equations \eqref{Boundary1} and \eqref{Boundary2} as comprising \emph{one side} of the usual Israel junction conditions across a singular hypersurface embedded in the underlying manifold \cite{Israel,Shtanov,ClockworkCosmo}.

Lastly, since one has in the bulk that $\nabla_N G^{MN} = \nabla_N T^{MN} = 0$, by the Gauss-Codazzi relations (see e.g. \cite{wald}) the brane energy-momentum tensors are also conserved:
\begin{equation}
    D_\mu \Theta^{\mu\nu} = 0 \; ;
\end{equation}
this implies that in the deconstructed theory the Bianchi constraint takes the form of Eq. \eqref{NewBianchiConstraint} for every $W$-tensor (as opposed to the sum in Eq. \eqref{W sum}).

The solutions of the RS1 system outlined above are very well-studied (see e.g. \cite{Langlois_review,Carsten_review} and references therein); we are going to focus on the realm of cosmology as a concrete example to show that our deconstruction procedure can recover 5-dimensional GR. Cosmological solutions of RS1 are known \cite{Full_EFEs_brane,Maeda2000,GeometryofBraneWorld}; they can always be written in so-called `Gaussian-normal' coordinates in the following manner:
\begin{equation}\label{RS1_continuum}
    \dd{s}^2 = -c^2(t,y) \dd{t}^2 + a^2(t,y) \gamma_{ij}\dd{x}^i\dd{x}^j + b^2(t,y) \dd{y}^2 \; ,
\end{equation}
where $\gamma_{ij}$ is a maximally symmetric spatial metric in 3-dimensions (with spatial curvature $k=-1,0,1$). 

This form of the metrics is useful because it is written in exactly the 4+1 ADM variables with which we did our deconstructing. Indeed, comparing Eq. \eqref{RS1_continuum} with Eq. \eqref{ADM} tells us that we are working in a gauge where the lapse is $\mathcal{N}=b(t,y)$, the shift is $N^\mu=0$, and the metric of the 4-dimensional hypersurfaces $\Sigma_y$ is:
\begin{equation}\label{5D cosmology}
    g_{\mu\nu}\dd x^\mu \dd x^\nu = -c^2(t,y)\dd t^2 + a^2(t,y)\gamma_{ij}\dd x^i \dd x^j \; ,
\end{equation}
which is just the usual FLRW metric with another lapse function in the time direction. For simplicity, we will from now on assume 0 spatial curvature in these induced metrics, so that $\gamma_{ij}=\eta_{ij}$. We will also take the matter on the branes to be of perfect fluid form, so that:
\begin{equation}\label{perfect_fluid_Tmunu}
    \Theta^{(i)\mu}_{\;\;\;\;\;\nu} = \text{diag}\left(- (\rho_i + \sigma_i), \;p_i-\sigma_i, \;p_i-\sigma_i, \;p_i-\sigma_i\right) \; ,
\end{equation}
where $\rho_i$ and $p_i$ are the energy density and pressure of the fluid.

Inserting the ansatz \eqref{RS1_continuum} into the 5-dimensional Einstein equations, the bulk field equations are found to be \cite{Carsten_review,Langlois_review}:
\begin{align}\label{G00}
    G^{0}_{\;0} &= \frac{3}{c^2}\left(\frac{\dot{a}^2}{a^2}+\frac{\dot{a}}{a}\frac{\dot{b}}{b}\right) + \frac{3}{b^2} \left(\frac{a'}{a}\frac{b'}{b} -\frac{a''}{a}-\frac{a^{\prime2}}{a^2}\right) = \frac{2\Lambda_5}{M_{(5)}^3} \; ,
    \\
    \begin{split}\label{Gij}
    G^{i}_{\;j} &= \frac{\delta^i_j}{c^2} \left(-\frac{\dot{a}^2}{a^2}+2\frac{\dot{a}}{a}\frac{\dot{c}}{c}-2\frac{\dot{a}}{a}\frac{\dot{b}}{b}+\frac{\dot{b}}{b}\frac{\dot{c}}{c}-2\frac{\ddot{a}}{a}-\frac{\ddot{b}}{b}\right) \\&+ \frac{\delta^i_j}{b^2} \left(\frac{a^{\prime2}}{a^2}-2\frac{a^\prime}{a}\frac{b^\prime}{b}+2\frac{a'}{a}\frac{c'}{c}-\frac{b'}{b}\frac{c'}{c}+2\frac{a''}{a}+\frac{c''}{c}\right) = -\frac{2\Lambda_5}{M_{(5)}^3}\delta^i_j \; ,
    \end{split}
    \\
    G^{5}_{\;0} &= \frac{3}{b^2}\left( \frac{\dot{a}}{a}\frac{c^\prime}{c}+\frac{a^\prime}{a}\frac{\dot{b}}{b}-\frac{\dot{a}'}{a}\right) = 0\label{G05} \; ,
    \\
    G^{5}_{\;5} &= \frac{3}{b^2}\left( \frac{a^{\prime2}}{a^2}+\frac{a'}{a}\frac{c'}{c}\right) + \frac{3}{c^2}\left(\frac{\dot{a}}{a}\frac{\dot{c}}{c}-\frac{\dot{a}^2}{a^2}-\frac{\ddot{a}}{a}\right) = -\frac{2\Lambda_5}{M_{(5)}^3}\label{G55} \; ,
\end{align}
while the boundary equations/junction conditions are:
\begin{align}
    K^0_{\;0} &\implies \left.3M_{(5)}^3\frac{a'}{ab}\right|_{0,L} = \mp(\rho_{0,L} + \sigma_{0,L}) \; ,\label{K00}
    \\
    K^i_{\;j} &\implies \left. M_{(5)}^3\left(2\frac{a'}{ab} + \frac{c'}{cb}\right)\right|_{0,L} = \pm( p_{0,L} - \sigma_{0,L}) \; ,\label{Kij}
\end{align}
where the dots represent time derivatives and primes represent $y$-derivatives.

\subsubsection{Deconstructed theory}

\begin{figure}[h!]
    \centering
    \includegraphics[width=\textwidth]{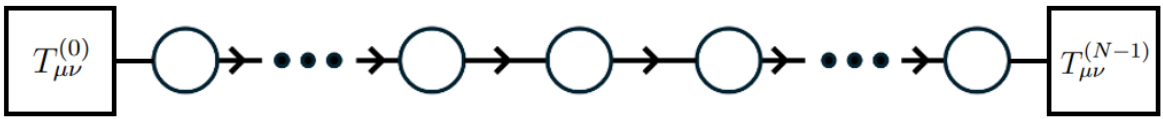}
    \caption{Theory graph for the deconstructed STMG theory that corresponds to the RS1 model in its continuum limit. Matter is coupled to the metrics at either end of the chain of interactions, which is analogous in the continuum to placing matter on end-of-the-world branes located at the two components of the extra dimensional boundary.}
    \label{fig:chain matter}
\end{figure}

We want to recover each of the 6 equations \eqref{G00}--\eqref{Kij} from our deconstructed STMG theory upon taking the continuum limit. The deconstructed theory that corresponds to pure GR in its continuum limit, as we have seen (c.f Eqs. \eqref{Tz tilde}--\eqref{Tiii tilde} when $\alpha_1=\alpha_2=\alpha_3=0$), involves chain-type interactions with the following choice for the non-vanishing interaction coefficients:
\begin{align}\label{NewBetas}
    \beta_{m}^{(i,i+1)}(\bm{\mathcal{N}}) = \frac{2}{\mathcal{N}_i+\mathcal{N}_{i+1}}\bar{\beta}_m \; ,
\end{align}
where:
\begin{align}
    \bar{\beta}_0 &= -\frac{6M_{(5)}^3}{\dy} \; ,
    \\
    \bar{\beta}_1 &= \frac{3M_{(5)}^3}{\dy} \; ,
    \\
    \bar{\beta}_2 &= -\frac{M_{(5)}^3}{\dy} \; ,
    \\
    \bar{\beta}_3=\bar{\beta}_4 &= 0 \; .
\end{align}
Putting matter on the branes located at the extra dimensional boundaries translates into the language of the deconstructed theory as having a matter sector that contains only non-vanishing $T^{(0)}_{\mu\nu}$ and $T^{(N-1)}_{\mu\nu}$, as displayed in figure \ref{fig:chain matter}.

Comparison with the continuum expression \eqref{RS1_continuum} for the 5-dimensional metric means that we need to take the 4-dimensional multi-gravity metrics to be:
\begin{equation}\label{FRW}
    \gi{i}\dd x^\mu \dd x^\nu = -c_i^2(t)\dd t^2 + a^2_i(t) \eta_{ij}\dd x^i \dd x^j \; ,
\end{equation}
and the additional scalar fields to be:
\begin{equation}\label{FRW lapse}
    \mathcal{N}_i = b_i(t) \; ,
\end{equation}
where, of course, $a_i(t)=a(t,y_i)$, $b_i(t)=b(t,y_i)$ and $c_i(t)=c(t,y_i)$.

If our deconstruction procedure is correct, upon substituting these ansatze into the STMG field equations, \eqref{NewFieldEqs}, the equations for $G^{(i)0}_{\;\;\;\;\;0}$ and $G^{(i)j}_{\;\;\;\;\;k}$ should respectively become Eqs. \eqref{G00} and \eqref{Gij} for all sites living in the bulk, or \eqref{K00} and \eqref{Kij} for each of the two sites on the boundaries, once $\dy$ is sent to 0. In the same limit, the Bianchi constraint, Eq. \eqref{NewBianchiConstraint}, should become the 5-dimensional momentum constraint \eqref{G05}, and the new scalar equation, Eq. \eqref{ScalarConstraint}, should become the 5-dimensional Hamiltonian constraint, Eq. \eqref{G55}.

In appendix \ref{App:brane cosmo}, we go through each of these calculations in turn, explicitly deriving the Friedmann equations for STMG when one takes the cosmological ansatze \eqref{FRW} and \eqref{FRW lapse} for the metrics and scalar fields, as well as the explicit form of the scalar equation and Bianchi constraint. Upon taking the continuum limit of these equations, we demonstrate that all of the above statements are indeed true, so that one recovers the dynamics and constraints of brane cosmology \emph{in their entirety} from the STMG theory with interaction coefficients given by \eqref{NewBetas} (the choice corresponding to pure GR in the continuum). In particular, the new scalar equations, which were missing in standard multi-gravity but are present in STMG, encode the Hamiltonian constraint in the extra dimensional theory; consequently, one hopes that the problems of standard multi-gravity that we described in section \ref{sec:fixing lapse}, related to the missing lapse function, will be resolved by the presence of the additional scalar fields and their corresponding field equations in our new theory. 

As mentioned back in section \ref{sec:fixing lapse}, we will save investigating whether this is indeed the case for future work, but we now at least possess the framework within which one should begin these investigations: it seems that STMG -- not standard multi-gravity -- really is the 4-dimensional gravitational theory that arises when one deconstructs GR in 5-dimensions (subject to confirming that it remains valid all the way up to the 5-dimensional Planck scale, and not only up to the lower strong coupling scale, $\Lambda_{\text{SC}}$, defined in section \ref{sec:fixing lapse}).

\section{Scalar-tensor multi-gravity}\label{Sec:MMST}

Despite the discussions at the end of the last section, standard multi-gravity, when considered on its own without reference to dimensional deconstruction, is a perfectly viable 4-dimensional theory of modified gravity. It can, depending on the masses of its various multi-gravitons, satisfy the usual tests of GR \cite{Solar_sys_tests,Recovering_GR,Vainshtein_recovery,Graviton_mass_bounds,heavy_spin2_DM}, as well as provide interesting new takes on various long-standing problems existing at the interface between gravity and particle physics (e.g. the Higgs hierarchy problem \cite{ClockworkGrav,ClockworkCosmo}, the nature of dark matter \cite{heavy_spin2_DM,DM_multigrav,bigravity_DM,Oscillating_DM,spin2portal,me_and_Lucien} etc.). Bigravity, the simplest of the multi-metric theories, has been shown to admit perfectly viable cosmologies \cite{viable_cosmologies,analytical_constraints_bigravity,observational_constraints_bigravity,combining_cosmo_local_bounds_bigravity}. Beyond this, there are suggestions that certain phenomena more typically thought of in the context of extra dimensions -- most notably the Gregory-Laflamme instability that afflicts black string configurations in higher dimensional gravity \cite{GL_instability,AdS_GL_instability,Charged_GL_instability} -- may actually have their roots in the massive spin-2 substructure that we have uncovered throughout section \ref{Sec:new deconstruction} \cite{GL_instability_bigravity,Kerr_instability_bigravity,BHs_multigrav,BH2,BHs_review}. 

Consequently, it behoves us to consider STMG as a potentially interesting gravitational theory in its own right, away from the continuum limit, or even away from situations where such a limit can be sensibly defined. Since we have shown that STMG is better behaved than standard multi-gravity in the situations where it makes sense to talk about a continuum limit (i.e. when one has chain-type interactions between the metrics), we are motivated to study how these hybrid theories of multi-gravity and scalar fields may differ from standard multi-gravity purely at the level of the lower-dimensional theories themselves, without reference to deconstruction. Hence, we now come to generalise the theory we developed in section \ref{Sec:new deconstruction} to arbitrary dimension and interaction structures, as well as allowing for more generic couplings of the scalars in the kinetic sector. We will also determine some of its simple solutions, which we will show differ from the corresponding solutions in standard multi-gravity precisely as a consequence of the new scalar equations.

\subsection{Metric and vielbein formulations in arbitrary dimension}

To generalise the STMG action \eqref{NewActionMetric}, we want to write it in arbitrary dimension, as well as allow for more generic couplings to the scalar fields (which in this section we give the more conventional name $\phi_i$) in the kinetic sector, beyond the Brans-Dicke type couplings of section \ref{Sec:new deconstruction}. For aesthetic purposes, we will also reabsorb the cosmological constant contribution, $\Lambda_4$, into the definition of $\beta_m^{(i,j)}$. Therefore, throughout this section we will be considering the following theory, written in the metric formulation as:
\begin{align}
    I &= I_K + I_V + I_M\left[\psi_i; g_{(i)}\right]\label{STMG action}
    \\
    I_K &= \sum_{i=0}^{N-1} \int \dd[D]x\sqrt{-\det g_{(i)}} \, \left[\frac{F_i(\phi_i)}{2}R_{(i)} - \frac{G_i(\phi_i)}{2}g^{(i)\mu\nu}\partial_\mu\phi_i\partial_\nu\phi_i - U_i(\phi_i)\right]
    \\
    I_V &= -\sum_{i,j} \int \dd[D]x \sqrt{-\det g_{(i)}} \sum_{m=0}^{D} \beta_m^{(i,j)}\left(\bm{\phi}\right) e_m(S_{i\rightarrow j})  \; ,
\end{align}
where the notation for the matter action $I_M\left[\psi_i; g_{(i)}\right]$ means that the matter fields, $\psi_i$, are minimally coupled to their corresponding metrics $\gi{i}$. Theories that have $F_i(\phi_i)=M_i^{D-2}\phi_i$ and $G_i(\phi_i)=M_i^{D-2}\omega_i/\phi_i$ form the Brans-Dicke subclass; if all $\omega_i=0$\footnote{In this case, the scalar potentials $U_i(\phi_i)$ may be reabsorbed into the definitions of $\beta_m^{(i,j)}(\bm{\phi})$ -- specifically $\beta_0$ -- to recover precisely the $\omega=0$ STMG action \eqref{NewActionMetric}, although it is natural to want to explicitly separate out the scalar potentials from the parts of the $\beta_m^{(i,j)}$ coefficients that correspond to genuine non-minimal interactions between the scalars and the metrics.} then one finds precisely the special set of STMG theories that arose from dimensional deconstruction, with which we are now familiar from section \ref{Sec:new deconstruction}. Recall that, owing to the arguments of \cite{MVMG,Generalised_massive_grav,Ghost_freedom_quadratic_bigravity}, the field content of the theory defined by Eq. \eqref{STMG action} should comprise a single massless spin-2 field, $N-1$ massive spin-2 fields, and $N$ scalar fields, with no ghosts (provided that there are no interaction cycles in the theory graph).

We have written this action in unitary gauge for the Stückelberg fields, but as always we are free to include them by making the replacement $S_{i\rightarrow j}\rightarrow \tilde{S}_{i\rightarrow j}$, given by Eq. \eqref{Stuckelberg metric}. We also note that the action is written in the so-called \emph{Jordan frame}, where the scalars couple directly to the Ricci scalar but not directly to $I_M$. In ordinary scalar-tensor theories, one may always perform scalar-dependent conformal transformations on the metrics to bring the action into the \emph{Einstein frame}, where the scalars instead couple directly to matter but only minimally to gravity (see e.g. \cite{ST_book}). We are going to work in the Jordan frame for the remainder of this work since it facilitates comparison with what we did throughout section \ref{Sec:new deconstruction}, as the Jordan frame scalars correspond to the discretised lapse, but we include the conversion to the Einstein frame version of the theory in appendix \ref{app:Einstein frame}. The generic structure of the multi-metric potential remains the same in either frame; the only difference between them is the explicit manner in which the scalar potentials $U_i(\phi_i)$ and interaction coefficients $\beta_m^{(i,j)}(\bm{\phi})$ depend on $\phi_i$. For example, the bi-metric Starobinsky model we mentioned in section \ref{Sec:new deconstruction} as an example STMG theory that has already been studied in an inflationary context, has an action which takes the form of the Einstein frame version of Eq. \eqref{STMG action}, with schematically $\beta_m(\bm{\phi})\sim e^{(D-m)\phi_1} e^{m\phi_2}\bar{\beta}_m$.

The vielbein form of the STMG action \eqref{STMG action} is given by:
\begin{align}
    I &= I_K + I_V + I_M\left[\psi_i; e^{(i)}\right]\label{STMG action vielbein}
    \\
    I_K &= \sum_{i=0}^{N-1} \int_{\mathcal{M}_D} \left[\frac{F_i(\phi_i)}{2}R^{(i)}_{ab} \wedge \hodge^{(i)} e^{(i)ab} - \frac{G_i(\phi_i)}{2}\dd\phi_i\wedge\hodge^{(i)}\dd\phi_i - U_i(\phi_i)\hodge^{(i)}1\right]
    \\
    I_V &= -\sum_{i_1\hdots i_D=0}^{N-1} \int_{\mathcal{M}_D} \varepsilon_{a_1\hdots a_D} T_{i_1\hdots i_D}(\bm{\phi}) e^{(i_1)a_1}\wedge\hdots\wedge e^{(i_D)a_D} \; ,
\end{align}
where $\dd$ denotes the exterior derivative. By the same discussion we had in section \ref{Sec:vielbein}, the metric and vielbein formulations are equivalent only for pairwise interactions where the Deser-van Nieuwenhuisen symmetric vielbein condition \eqref{SymmetricVierbeinCondition} holds. As we saw back then, the vielbein formalism is generically less restrictive than the metric formalism; non-pairwise-interacting multi-vielbein theories, where the $\Tcoeffs$ factorise into $T_{i_1}\hdots T_{i_D}$, are also ghost free in standard multi-gravity \cite{beyond_pairwise_couplings,mass_spectrum_multivielbein}. Presumably, this newer class of interactions remains ghost free in STMG too.

As we have done throughout this paper, we will consider only the pairwise-interacting theories where the metric and vielbein formulations are interchangeable, in which case, the field equations become:
\begin{equation}\label{STMG field eqs}
\boxed{
    F_i(\phi_i) G^{(i)}_{\mu\nu} + W^{(i)}_{\mu\nu}(\bm{\phi}) = T^{(i, m)}_{\mu\nu} + T^{(i,\phi)}_{\mu\nu} }\; ,
\end{equation}
where $T^{(i,m)}_{\mu\nu}$ is the energy-momentum tensor for the matter sector residing within $I_M$, while the energy-momentum tensor for the scalars $\phi_i$ is:
\begin{equation}
\begin{split}
    T^{(i,\phi)}_{\mu\nu} &= F_i'(\phi_i)\left(\nabla^{(i)}_\mu\nabla^{(i)}_\nu\phi_i-\gi{i}\Box^{(i)}\phi_i\right)  - \gi{i} U_i(\phi_i)
    \\
    &+ \left[\left(G_i(\phi_i)+F''_i(\phi_i)\right)\left(\partial_\mu\phi_i \partial_\nu\phi_i -\frac12 \left(G_i(\phi_i)+2F''_i(\phi_i)\right)\gi{i}\partial_\lambda\phi_i\partial^\lambda\phi_i\right)\right] \; ,
\end{split}
\end{equation}
with primes now denoting derivatives with respect to $\phi_i$. In the $\omega=0$ Brans-Dicke case where $F_i(\phi_i)=M_i^{D-2}\phi_i$ and $G_i(\phi_i)=0$, only the first term in this expression survives, while the potential part is simply absorbed into the $\beta_m^{(i,j)}(\bm{\phi})$ coefficients of the $W$-tensors, thus recovering Eq. \eqref{NewFieldEqs}.

The $W$-tensors, in metric terms, take the form:
\begin{equation}\label{STMG W}
    \Wi{i}(\bm{\phi}) = \sum_j \sum_{m=0}^D(-1)^m\beta_m^{(i,j)}(\bm{\phi})Y_{(m)\nu}^\mu(S_{i\rightarrow j})
    +\sum_k\sum_{m=0}^D (-1)^m\beta_{D-m}^{(k,i)}(\bm{\phi})Y_{(m)\nu}^\mu (S_{i\rightarrow k}) \; ,
\end{equation}
where as usual $j$ denote positively oriented interactions and $k$ denote negatively oriented interactions with respect to $\gi{i}$. The corresponding vielbein form expression is:
\begin{equation}
    \Wi{i}(\bm{\phi}) = D!\vbein{i}{\nu}{a}\ivbein{i}{\mu}{[a}\ivbein{i}{\lambda_1}{b_1}\hdots\ivbein{i}{\lambda_{D-1}}{b_{D-1}]} \sum_{j_1\hdots j_{D-1}} \mathcal{P}(i) T_{ij_1\hdots j_{D-1}}(\bm{\phi}) \vbein{j_1}{\lambda_1}{b_1}\hdots\vbein{j_{D-1}}{\lambda_{D-1}}{b_{D-1}} \; ;
\end{equation}
of course, for pairwise interactions, the above two definitions can be shown to be equivalent by virtue of Eqs. \eqref{betas1} and \eqref{betas2}, as was the case in standard multi-gravity.

Assuming general covariance of the complete matter sector across all sites (i.e. only that which is contained within $I_M$, not the scalars), the divergences of the $W$-tensors must satisfy:
\begin{equation}
    \sum_{i=0}^{N-1} \sqrt{-\det g_{(i)}}\left[\nabla^{(i)}_{\mu}\Wi{i}(\bm{\phi}) + F_i'(\phi_i)\partial_\mu\phi_i G^{(i)\mu}_{\;\;\;\;\;\nu} - \nabla^{(i)}_\mu T^{(i,\phi)\mu}_{\;\;\;\;\;\;\;\;\;\nu}\right] = 0 \; .
\end{equation}
If one further assumes that the matter energy-momentum tensors are covariantly conserved on each individual site (which is always true if there is only a single matter coupling anyway) i.e.
\begin{equation}
    \nabla^{(i)\mu} T^{(i,m)}_{\mu\nu} =0 \; ,
\end{equation}
then this becomes the modified Bianchi constraint:
\begin{equation}\label{STMG Bianchi}
\boxed{
    \nabla^{(i)}_{\mu}\Wi{i}(\bm{\phi}) = \nabla^{(i)}_\mu T^{(i,\phi)\mu}_{\;\;\;\;\;\;\;\;\;\nu} -F_i'(\phi_i)\partial_\mu\phi_i G^{(i)\mu}_{\;\;\;\;\;\nu}}\; .
\end{equation}
Again, when $F_i(\phi_i)=M_i^{D-2}\phi_i$ and $G_i(\phi_i)=0$, the divergence of the surviving term in the scalar energy-momentum tensor combines with the Einstein tensor part to retrieve the familiar form of the Bianchi constraint from section \ref{Sec:new deconstruction} (c.f. Eq. \eqref{NewBianchiConstraint}).

Lastly, the equations of motion for the scalar fields $\phi_i$ read as follows:
\begin{equation}\label{STMG scalar eq}
\boxed{
    G_i(\phi_i)\Box^{(i)}\phi_i + \frac{F_i'(\phi_i)}{2}R_{(i)} + \frac{G_i'(\phi_i)}{2}g^{(i)\mu\nu}\partial_\mu\phi_i\partial_\nu\phi_i = U_i'(\phi_i) + \mathbb{X}_i(\bm{\phi})
    } \; ,
\end{equation}
where in metric form the new term on the right hand side is:
\begin{equation}\label{metric X}
    \mathbb{X}_i(\bm{\phi}) = \sum_j\sum_{m=0}^D \frac{\partial\beta_m^{(i,j)}}{\partial\phi_i}e_m(S_{i\rightarrow j}) + \sum_k\sum_{m=0}^D \frac{\partial\beta_{D-m}^{(k,i)}}{\partial\phi_i}e_m(S_{i\rightarrow k}) \; ,
\end{equation}
and in vielbein form it is:
\begin{equation}
    \mathbb{X}_i(\bm{\phi}) = D! \ivbein{i}{\mu_1}{[a_1}\hdots\ivbein{i}{\mu_D}{a_D]}\sum_{j_1\hdots j_D} \frac{\partial T_{j_1\hdots j_D}}{\partial\phi_i}\vbein{j_1}{\mu_1}{a_1}\hdots\vbein{j_D}{\mu_D}{a_D} \; .
\end{equation}
Again, these two definitions of $\mathbb{X}_i$ are equivalent for pairwise interactions, and when $F_i(\phi_i)=M_i^{D-2}\phi_i$ and $G_i(\phi_i)=0$, Eq. \eqref{STMG scalar eq} reduces to Eq. \eqref{ScalarConstraint} from section \ref{Sec:new deconstruction} (after absorbing the potential part into the interaction coefficients as before). 

As alluded to in section \ref{Sec:new deconstruction}, one may explicitly extract the dynamics of the scalar fields $\phi_i$ from Eq. \eqref{STMG scalar eq} by taking the trace of Eqs. \eqref{STMG field eqs} and substituting in for the Ricci scalar. This trace reads:
\begin{equation}
    \begin{split}
        \frac{2-D}{2}F_iR_{(i)}+W_{(i)} = T_{(i,m)} + (\partial\phi_i)^2\left[\frac{2-D}{2}G_i+(1-D)F''_i\right]+F_i'(1-D)\Box^{(i)}\phi_i -DU_i \; ,
    \end{split}
\end{equation}
where $T_{(i,m)}=g^{(i)\mu\nu}T^{(i,m)}_{\mu\nu}$ is the trace of the $i$-th matter energy-momentum tensor, and $W_{(i)}=g^{(i)\mu\nu}W^{(i)}_{\mu\nu}$ is the trace of the $i$-th $W$-tensor, which owing to Eq. \eqref{Trace Y} is:
\begin{equation}
    W_{(i)}(\bm{\phi}) = \sum_j\sum_{m=0}^D \beta_m^{(i,j)}(\bm{\phi})(D-m)e_m(S_{i\rightarrow j}) + \sum_k\sum_{m=0}^D \beta_{D-m}^{(k,i)}(\bm{\phi})(D-m)e_m(S_{i\rightarrow k}) \; .
\end{equation}
Rearranging and substituting in for $R_{(i)}$ in Eq. \eqref{STMG scalar eq} leads to:
\begin{equation}\label{STMG scalar dynamics}
\boxed{
\begin{aligned}
    \left(G_i + \frac{D-1}{D-2}\frac{F^{\prime2}_i}{F_i}\right)\Box^{(i)}\phi_i = &\frac{F_i'}{F_i}\frac{T_{(i,m)}}{D-2} + U'_i - U_i\frac{F_i'}{F_i}\frac{D}{D-2} 
    \\
    &- (\partial\phi_i)^2\left(\frac{G_i}{2}\frac{F'_i}{F_i}+\frac{D-1}{D-2}\frac{F_i'}{F_i}F_i'' + \frac{G_i'}{2}\right)
    \\
    &+ \mathbb{X}_i - \frac{F_i'}{F_i}\frac{W_{(i)}}{D-2} 
\end{aligned}
} \; ,
\end{equation}
so that all derivatives now act exclusively on the scalars and not on the metrics. In the Brans-Dicke case where $F_i(\phi_i) = M_i^{D-2}\phi_i$ and $G_i(\phi_i)=M_i^{D-2}\omega_i/\phi_i$, the coefficient of $(\partial\phi_i)^2$ vanishes. If the coefficient of $\Box^{(i)}\phi_i$ also vanishes, which happens in the Brans-Dicke case when $\omega_i=-\frac{D-1}{D-2}$, then the scalars are actually non-propagating, but for any other value they are dynamical. Lastly, we note that the part of these scalar equations that depends on the non-minimal interactions with the metrics can be written:
\begin{equation}
    \begin{split}
        \mathbb{X}_i(\bm{\phi}) - \frac{F_i'(\phi_i)}{F_i(\phi_i)}\frac{W_{(i)}(\bm{\phi})}{D-2} &= \sum_j\sum_{m=0}^D \left[\frac{\partial\beta_m^{(i,j)}}{\partial\phi_i}-\frac{F_i'}{F_i}\frac{D-m}{D-2}\beta_m^{(i,j)}\right]e_m(S_{i\rightarrow j}) 
        \\
        &+ \sum_k\sum_{m=0}^D \left[\frac{\partial\beta_{D-m}^{(k,i)}}{\partial\phi_i}-\frac{F_i'}{F_i}\frac{D-m}{D-2}\beta_{D-m}^{(k,i)}\right]e_m(S_{i\rightarrow k}) \; .
    \end{split}
\end{equation}

Eqs. \eqref{STMG field eqs}, \eqref{STMG Bianchi} and \eqref{STMG scalar dynamics} comprise the full set of equations and constraints that define a given STMG system in the Jordan frame\footnote{All of these field equations look much simpler in the Einstein frame, which we include in appendix \ref{app:Einstein frame}. There, the intuition of the scalars as corresponding to the deconstructed lapse function is sacrificed in favour of operational comfort (but of course, if we are working with STMG away from any sort of continuum limit, this intuition is not worth much anyway).}. If $F_i(\phi_i) = M_i^{D-2}\phi_i$ and $G_i(\phi_i)=0$, they are equivalent to the equations of the $\omega=0$ Brans-Dicke type theory from section \ref{Sec:new deconstruction}, and if all of the scalars are fixed to $\phi_i=1$, they are equivalent to the equations of standard multi-gravity from section \ref{Sec:review}, plus the additional scalar equations \eqref{STMG scalar dynamics}. Let us now see if we can find some simple solutions of this system to compare against what happens in standard multi-gravity. We are going to return to the $\omega=0$ Brans-Dicke case where $F_i(\phi_i) = M_i^{D-2}\phi_i$ and $G_i(\phi_i)=0$, both for simplicity, and to tie in better with everything we did in the previous sections.

\subsection{Comparison with standard multi-gravity: what do the extra scalars do?}

Let us work in vacuum, where $I_M=0$, for simplicity. In standard multi-gravity, the simplest vacuum solutions one may construct are the \emph{proportional solutions}, where, as the name suggests, all of the various metrics are proportional to one another \cite{prop_bg_multigrav,consistent_spin2}:
\begin{equation}\label{propmetrics}
    \gi{i} = a_i^2 g^{(\Lambda)}_{\mu\nu} \; .
\end{equation}
Here, $g^{(\Lambda)}$ is the metric of some Einstein space with effective cosmological constant $\Lambda$ (which may be 0, but it has to be the same on all sites if the theory is to admit rotating black hole solutions \cite{BHs_multigrav}) and the $a_i$ are conformal factors that the Bianchi constraint forces to be constant \cite{consistent_spin2}. With this ansatz, the standard multi-gravity vacuum equations $M_i^{D-2}G^{(i)}_{\mu\nu}+W^{(i)}_{\mu\nu}=0$ take the form \cite{BH2}:
\begin{equation}\label{propsols}
    \frac{\Lambda M_i^{D-2}}{a_i^2} = \sum_j W_{i,j}^{(+)} +\sum_k W_{i,k}^{(-)}  \;\;\;\;\;\forall\, i \; ,
\end{equation}
where we define:
\begin{align}
    W_{i,j}^{(+)} &= \sum_{m=0}^{D} \beta_m^{(i,j)} \binom{D-1}{m}\left(\frac{a_j}{a_i}\right)^m \; ,
    \\
    W_{i,k}^{(-)} &= \sum_{m=0}^{D} \beta_{D-m}^{(k,i)} \binom{D-1}{m}\left(\frac{a_k}{a_i}\right)^m \; .
\end{align}
After fixing one of the $a_i$ via coordinate rescaling, Eqs. \eqref{propsols} comprise $N$ algebraic, nonlinear simultaneous equations that may be solved for $\Lambda$ and the remaining $N-1$ conformal factors, the physical solutions being those with real $\Lambda$ and $a_i$. In this way, standard multi-gravity naturally admits de Sitter, anti-de Sitter and Minkowski vacua, where the interactions between metrics manifest themselves as an effective cosmological constant.

In hindsight, the fact that solutions with non-zero $\Lambda$ exist in standard multi-gravity should have already raised alarm bells about its ability to arise from dimensional deconstruction; taken with chain-type interactions, in the (ill-defined) continuum limit, the ansatz \eqref{propmetrics} corresponds to the 5-dimensional geometry:
\begin{equation}
    \dd s^2 = a^2(y)g^{(\Lambda)}_{\mu\nu}\dd x^\mu \dd x^\nu +\dd y^2 \; ,
\end{equation}
which one can check is \emph{not} a solution of 5-dimensional GR on a bounded manifold unless $\Lambda=0$. However, if we allow for an arbitrary lapse $\phi(y)$ in the $y$-direction, such that the 5-dimensional geometry is:
\begin{equation}
    \dd s^2 = a^2(y)g^{(\Lambda)}_{\mu\nu}\dd x^\mu \dd x^\nu +\phi^2(y)\dd y^2 \; ,
\end{equation}
then solutions with non-vanishing $\Lambda$ \emph{do} exist. As a result, we should find that in STMG -- which we claim can directly descend from 5-dimensions for well-chosen $\beta_m^{(i,j)}(\bm{\phi})$ -- there are no solutions where both $\Lambda\neq 0$ and $\phi_i=1$ simultaneously, but $\Lambda\neq0$ solutions emerge once the $\phi_i$ are allowed to be arbitrary. Let us see how this works in practice.

\begin{figure}[h!]
\centering
    \includegraphics[width=0.4\textwidth]{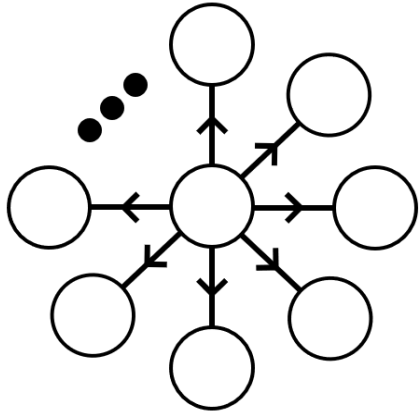}
    \caption{Theory graph for a star-type interaction.}
    \label{fig:star}
\end{figure}

For this calculation, it proves simplest to work with star-type interactions, where all multi-gravity metrics couple to some central metric but not to each other, as in figure \ref{fig:star}. However, one can show (it is just a bit fiddlier to do so) that the end result is the same for chain-type interactions too, so everything we will present below holds for arbitrary interaction structures, which can always be built by stringing together combinations of stars and chains.

Consider the STMG equations and constraints with the ansatz \eqref{propmetrics} for the metrics. If the $a_i$ are constant, then the Bianchi constraint \eqref{STMG Bianchi} immediately forces the $\phi_i$ to be constant too. If we try to assert that all $\phi_i=1$, then as we discussed, the metric field equations are the same as those in standard multi-gravity, \eqref{propsols}. However, we now also have the scalar equations \eqref{STMG scalar eq}, which were not present before; they read:
\begin{equation}\label{propscalarconstraints}
    \frac{D\Lambda}{D-2}\frac{M_i^{D-2}}{a_i^2} + \sum_j V_{i,j}^{(+)} + \sum_k V_{i,k}^{(-)} = 0\; ,
\end{equation}
where we have defined:
\begin{align}
    V_{i,j}^{(+)} &= -\sum_{m=0}^D \frac{\partial\beta_m^{(i,j)}}{\partial\phi_i} \binom{D}{m}\left(\frac{a_j}{a_i}\right)^m \; ,
    \\
    V_{i,k}^{(-)} &= -\sum_{m=0}^D \frac{\partial\beta_{D-m}^{(k,i)}}{\partial\phi_i} \binom{D}{m}\left(\frac{a_k}{a_i}\right)^m \; .
\end{align}
If one assumes that the $\beta_m^{(i,j)}(\bm{\phi})$ can be written as $f_{i,j}(\bm{\phi})\bar{\beta}_m$, and that the functions $f_{i,j}$ depend symmetrically on $\phi_i$ and $\phi_j$ in the sense that $\partial f/\partial\phi_i = \partial f/\partial\phi_j$, as was the case for our GR choice in Eq. \eqref{NewBetas}, then the $V$'s are related by:
\begin{equation}\label{V_relations}
    V_{i,j}^{(+)} = \left(\frac{a_j}{a_i}\right)^D V_{j,i}^{(-)} \; ,
\end{equation}
which is just a consequence of the identity \eqref{swap_orientation}. Similarly, if one assumes that $\left.\partial f/\partial\phi\right|_{\phi=1}=-1$, which is also the case for our GR choice of $\beta_m^{(i,j)}$, they are related to the $W_{i,j}^{(\pm)}$ by:
\begin{align}\label{V_W_relations}
    V_{i,j}^{(+)} &= W_{i,j}^{(+)} + \left(\frac{a_j}{a_i}\right)^D W_{j,i}^{(-)} \; ,
    \\
    V_{i,k}^{(-)} &= W_{i,k}^{(-)} + \left(\frac{a_k}{a_i}\right)^D W_{k,i}^{(+)} \; .\label{V_W_relations2}
\end{align}

For the star-type interaction, denoting the central metric as $\gi{0}$ and the outer metrics as $\gi{j}$, the scalar equations and metric field equations explicitly are:
\begin{multicols}{2}
\textbf{Central metric:}
    \begin{align}
        \frac{\Lambda M_0^{D-2}}{a_0^2} &= \sum_j W_{0,j}^{(+)}\label{W0}
        \\
        \frac{D\Lambda}{D-2}\frac{ M_0^{D-2}}{a_0^2} &=- \sum_{j} V_{0,j}^{(+)} \label{V0}
    \end{align}
\textbf{Outer metrics:}
    \begin{align}
        \frac{\Lambda M_0^{D-2}}{a_0^2} &= \left(\frac{a_j}{a_0}\right)^2\left(\frac{M_0}{M_j}\right)^{D-2} W_{j,0}^{(-)}\label{Wj}
        \\
        \frac{D\Lambda }{D-2}\frac{M_0^{D-2}}{a_0^2} &=- \left(\frac{a_j}{a_0}\right)^2\left(\frac{M_0}{M_j}\right)^{D-2} V_{j,0}^{(-)}  \label{Vj}
    \end{align}
\end{multicols}
\noindent where we have multiplied the equations for the outer metrics through by the factor $\left(a_j/a_0\right)^2\left(M_0/M_j\right)^{D-2}$ to make the left hand sides equal in all of the equations. One may then apply Eq. \eqref{V_relations} to the scalar equations to obtain:
\begin{equation}
    \left(\frac{M_0 a_0}{M_{j_1} a_{j_1}}\right)^{D-2} V_{0,j_1}^{(+)} = \left(\frac{M_0 a_0}{M_{j_2} a_{j_2}}\right)^{D-2} V_{0,j_2}^{(+)} = \hdots = \sum_j V_{0,j}^{(+)} \; .
\end{equation}
This implies:
\begin{equation}
    V_{0,j_k} = \left(\frac{M_{j_k} a_{j_k}}{M_{j_l} a_{j_l}}\right)^{D-2} V_{0,j_l}^{(+)} \; ,
\end{equation}
which we can substitute back in to show that the conformal factors must satisfy:
\begin{equation}\label{conformal factors}
    (M_0 a_0)^{D-2} = \sum_j (M_j a_j)^{D-2} \; .
\end{equation}
Turning our attention now to the field equations, using Eq. \eqref{V_W_relations} one can write:
\begin{equation}
    \sum_j W_{0,j}^{(+)} = \sum_j V_{0,j}^{(+)} - \sum_j \left(\frac{a_j}{a_i}\right)^D W_{j,0}^{(-)} \; .
\end{equation}
By Eq. \eqref{W0}, the first term becomes $\Lambda M_0^{D-2}/a_0^2$; by Eq. \eqref{V0}, the second term becomes $-(D\Lambda/(D-2)) M_0^{D-2}/a_0^2$; finally, by the combination of Eqs. \eqref{Wj} and \eqref{conformal factors}, the last term becomes $-\Lambda M_0^{D-2}/a_0^2$. 

Hence, one finds that the only way to satisfy both the metric and scalar field equations simultaneously, when all the scalar fields are fixed to $\phi_i=1$, requires:
\begin{align}
    \frac{\Lambda M_0^{D-2}}{a_0^2} &= -\frac{D\Lambda}{D-2}\frac{ M_0^{D-2}}{a_0^2} - \frac{\Lambda M_0^{D-2}}{a_0^2} \nonumber
    \\
    \implies 2\Lambda &= -\frac{D\Lambda}{D-2}\nonumber
    \\
    \implies \Lambda&=0 \; .
\end{align}
This is exactly what we wanted to show, and agrees with our intuition from dimensional deconstruction that there should not exist solutions with non-vanishing $\Lambda$ when the $\phi_i$ are all fixed to 1. Note that in the $\Lambda=0$ case, the field equations reduce down to just $W_{i,j}^{(\pm)}=0$; if these equations are satisfied, then the scalar equations are immediately satisfied as well, by virtue of Eqs. \eqref{V_W_relations} and \eqref{V_W_relations2}.

At first glance, this result might appear worrying: can it be that STMG does not admit any vacua with non-vanishing cosmological constant? If not, then the theory would fall down at the first hurdle, since it would not be able to describe the observed accelerated expansion of our universe\footnote{Well, unless any of the scalars manage to act as a quintessence field, but quintessence has enough of its own issues for us to not want to consider this possibility here.} \cite{Riess,Perlmutter}. Thankfully, we need not worry: as alluded to earlier, de Sitter and anti-de Sitter vacua do exist in the theory; it is just that we need to allow the $\phi_i$ to differ from site to site in order to find them. For generic $\phi_i$, the metric field equations \eqref{propsols} and scalar equations \eqref{propscalarconstraints}, \emph{when taken together}, comprise a set of $2N$ nonlinear simultaneous equations in the variables $\{\Lambda, a_i, \phi_i\}$. After fixing one of the conformal factors by choosing a coordinate system, there are $N-1$ free conformal factors $a_i$, $N$ scalar field VEVs $\phi_i$ and 1 effective cosmological constant $\Lambda$ for which we must solve; this makes $2N$ variables in total, so again the system is solvable in principle, the solutions being physical if all the variables turn out to be real. Thus, standard multi-gravity and STMG both possess Minkowski, de Sitter and anti-de Sitter vacua, but they differ in the manner in which they are constructed, thanks to the existence/non-existence of the scalar fields and their corresponding field equations. Note that this means all of the black hole solutions to standard multi-gravity, which we constructed in \cite{BHs_multigrav,BH2}, carry across naturally to STMG in this manner as well, being that they are also vacuum solutions.

\section{Discussion and conclusion}\label{sec:conclusion}

In this work, we sought to revisit and shed new light on an age-old question, namely: `What does gravitational physics in a theory with compact extra dimensions look like to a lower dimensional observer?' Folk wisdom from the time of Kaluza and Klein dictates that we should generically expect to see towers of massive spin-2 states in the spectrum of the corresponding 4-dimensional EFT; one can see this either following the standard KK framework, where one simply integrates out the extra dimension(s), or following the alternative prescription of dimensional deconstruction, where one discretises the extra dimension(s) on a lattice and treats the induced metrics on the various lattice sites as interacting tensor fields in the lower dimensional EFT (the two approaches are related to one another via discrete Fourier transforms of the vielbeins \cite{deconstructing_dims}). 

The deconstruction paradigm is extremely powerful, as it elucidates a direct link between pure gravity in higher dimensions and theories of massive and multi-gravity in lower dimensions, which have been extensively studied in recent years as viable alternatives to GR, after their ghost free formulations were discovered in 2010. However, extracting the precise form of the multi-gravity theory that arises from the deconstruction of, say, 5-dimensional GR, has historically proven difficult beyond perturbative level, owing to the inherent complexities baked into nonlinear theories of massive and multi-gravity. When this was originally attempted prior to the development of dRGT multi-gravity, the resulting 4-dimensional theory turned out to be ghostly and was therefore pathological. Still, even after the ghost free theories were developed, it was soon realised that they could not truly arise from the deconstruction of GR either, because they contain no counterpart of the higher dimensional lapse function. We discussed at length throughout the main body of this work why the missing lapse is problematic, and we outlined the various ways in which it prevents one from ever taking a well-defined continuum limit to recover 5-dimensional GR from standard 4-dimensional multi-gravity (for example, at the level of the deconstructed theory, it manifests as standard multi-gravity's (in)famously low strong coupling scale).

We have developed an improved deconstruction procedure that does not gauge fix the lapse before discretising the extra dimension, and we used it to demonstrate that the (tentatively) correct 4-dimensional theory that arises from the deconstruction of 5-dimensional GR is actually a hybrid of multi-gravity with a collection of scalar fields; we gave this theory the name `Scalar-Tensor Multi-Gravity', or STMG for short. STMG remains ghost free, as the structure of the spin-2 interactions is unchanged relative to standard multi-gravity (the scalars only appear implicitly in the multi-gravity interaction coefficients); the existence of the constraints that exorcise the ghost has already been explicitly proven for a simple bi-metric version of STMG studied recently in the context of inflation \cite{Ghost_freedom_quadratic_bigravity}. However, STMG is better behaved under deconstruction than standard multi-gravity because the field equations of the new scalars, which were missing in standard multi-gravity, encode the behaviour of the lapse function in the higher dimensional theory. Indeed, we showed explicitly, by means of the concrete example of Randall-Sundrum brane cosmology, that by taking the continuum limit of STMG, one is able to recover \emph{all} of the dynamical field equations, the momentum and Hamiltonian constraints, and the Israel junction conditions of 5-dimensional GR (with a compact extra dimension). However, the reason we still say that STMG is only \emph{tentatively} the correct deconstructed theory is that one still needs to confirm that it remains valid all the way up to the 5-dimensional Planck scale. To do this, one should investigate its decoupling limit to check that the theory does not become strongly coupled at an unacceptably low scale, a calculation that we save for future work.

Nevertheless, the development of this new theory, together with the recognition that even standard multi-gravity is a perfectly viable 4-dimensional EFT of modified gravity, motivated us to study theories involving multiple metrics/vielbeins \emph{and} scalars as potentially interesting EFTs in their own right, without reference to dimensional deconstruction. To this end, we generalised STMG to arbitrary dimension, to spin-2 interaction structures where there is no obvious notion of a continuum limit, and finally to generic scalar couplings in the kinetic sector. We derived all of the field equations of these STMG theories in both the Jordan and Einstein frames in complete generality, then focussed in on the $\omega=0$ Brans-Dicke subclass that arose from dimensional deconstruction to determine some simple vacuum solutions (the proportional solutions) and tie into our work from section \ref{Sec:new deconstruction}. We showed that, while these solutions are qualitatively the same as the corresponding solutions in standard multi-gravity, the manner in which they are quantitatively constructed differs and relies on the presence of the scalars and their associated field equations; we also showed why this makes sense from an extra dimensional standpoint, so everything seems to check out.

Going forward, we hope that this work will open a number of doors to interesting explorations. For example, immediately with STMG one is faced with the prospect of potentially developing new and viable models of screened modified gravity (see e.g. \cite{Brax_lectures}), where the new interactions between the scalar fields and spin-2 fields encoded implicitly in the $W$-tensors may have interesting phenomenological effects. There are potentially also nice dark matter scenarios to be uncovered involving the new scalar degrees of freedom -- indeed even standard multi-gravity can say a lot of interesting things in this regard \cite{heavy_spin2_DM,DM_multigrav,bigravity_DM,Oscillating_DM,spin2portal,me_and_Lucien}.

Even more enticing is that once one is armed with a consistent link between higher dimensional gravity and STMG in 4-dimensions, one may begin to rephrase questions more typically asked in the context of extra dimensions in a purely 4-dimensional language. Already we have seen signs that this may be a fruitful thing to do: we mentioned at the start of section \ref{Sec:MMST} that there are hints the Gregory-Laflamme instability plaguing higher dimensional black strings could potentially be a result of the black string's massive spin-2 substructure rather than its extra dimensional nature. This is because the GL instability persists in multi-gravity even when one is well away from the continuum limit, or has interaction structures where such a limit cannot even be sensibly defined \cite{BHs_multigrav,BH2}. For the chain-type interactions where it \emph{does} make sense to talk about taking the continuum limit, it would be extremely interesting to investigate whether any holographic interpretation exists for the boundary sites when the bulk is AdS, and to see whether such an interpretation survives the deconstruction process (or even better, if it exists away from the continuum limit, as was the case for the GL instability). Finally, many swampland conjectures are formulated with the implicit assumption of extra dimensions (e.g. those referring to string/KK towers of states) and it would be a good test of their robustness to see whether they still hold in multi-gravity outside of the continuum limit (where the towers of states are now finitely many massive gravitons, and there is no longer the implicit assumption that extra dimensions exist). 

We leave these interesting questions to future work, but we certainly intend to follow them up in the near future; we hope that we have also provided a comprehensive foundation upon which others may build if they too wish to join in the fun.

\section*{Acknowledgements}

I would like to thank Tasos Avgoustidis, Paul Saffin, Claudia de Rham, Zongzhe Du and Benjamin Muntz for useful discussions and comments on the manuscript. I am supported by a UK Science and Technology Facilities Council studentship, grant no. [ST/W507702/1]. For the purpose of open access, I have applied a Creative Commons Attribution (CC BY) licence to any Author Accepted Manuscript version arising.

\section*{Data access statement}
No new data were created or analysed in this study.

\appendix

\section{Explicit recovery of brane cosmology in the continuum limit of STMG}\label{App:brane cosmo}

In this appendix, we derive the form of the modified Friedmann equations, Bianchi constraint and scalar equations for the $\omega=0$ Brans-Dicke type STMG theory of section \ref{Sec:new deconstruction}, when one takes the FLRW ansatze for the metrics and scalars, given by Eqs. \eqref{FRW} and \eqref{FRW lapse}. We repeat these ansatze below for convenience:
\begin{align*}
    \gi{i}\dd x^\mu \dd x^\nu &= -c_i^2(t)\dd t^2 + a^2_i(t) \eta_{ij}\dd x^i \dd x^j \; ,
    \\
    \mathcal{N}_i &= b_i(t) \; .
\end{align*}
We will show, by explicitly taking the limit where $N\rightarrow\infty$ and $\dy\rightarrow0$ with $N\dy=L$ fixed, that the 4-dimensional equations of STMG -- Eqs. \eqref{NewFieldEqs}, \eqref{NewBianchiConstraint} and \eqref{ScalarConstraint} -- encode all of the 5-dimensional equations, constraints and junction conditions of Randall-Sundrum brane cosmology, given by Eqs. \eqref{G00}--\eqref{Kij}.

\subsection{Dynamical field equations}

First up are the RS1 dynamical equations i.e. the part of the continuum Einstein equations correpsonding to the dynamics of the hypersurface metrics $g_{\mu\nu}$. Recall that the field equations for the metrics in STMG are (c.f. Eq. \eqref{NewFieldEqs}):
\begin{equation}
    M_{(4)}^2 G^{(i)}_{\mu\nu} + 2\Lambda_4 \gi{i} + \frac{W^{(i)}_{\mu\nu}(\bm{b})}{b_i} = \frac{T^{(i)}_{\mu\nu}}{b_i} + \frac{M_{(4)}^2}{b_i} \left[\nabla^{(i)}_\mu\nabla^{(i)}_\nu b_i - \gi{i}\Box^{(i)}b_i\right] \; ,
\end{equation}
with the $W$-tensor defined by Eq. \eqref{NewW}, and where we have divided through by the scalars $b_i$ for convenience in what will follow. We want to show that the 00-component of these equations recovers Eq. \eqref{G00} upon taking the continuum limit, and that in the same limit the $jk$-components recover Eq. \eqref{Gij}.

With the cosmological ansatz \eqref{FRW} for the various metrics, the 4-dimensional Einstein tensor components are:
\begin{align}
    G^{(i)0}_{\;\;\;\;\;0} &= -3\left(\frac{\dot{a}_i}{a_i c_i}\right)^2 \; ,
    \\
    G^{(i)j}_{\;\;\;\;\;k} &= \frac{1}{c_i^2}\left(-\frac{\dot{a}^2_i}{a^2_i} - 2\frac{\ddot{a}_i}{a_i} + 2\frac{\dot{a}_i}{a_i}\frac{\dot{c}_i}{c_i}\right) \delta^j_k \; .
\end{align}
Also, the building-block matrices $S_{i\rightarrow j}$ take the simple form:
\begin{equation}
    S_{i\rightarrow j} = \text{diag}\left(\frac{c_j}{c_i}, \frac{a_j}{a_i}, \frac{a_j}{a_i}, \frac{a_j}{a_i}\right) \; ,
\end{equation}
which lead to the following non-vanishing $W$-tensor components \cite{ClockworkCosmo} (recalling that each site in the bulk has exactly one positively oriented interaction and one negatively oriented interaction -- see figure \ref{fig:chain matter}):
\begin{align}
    W^{(i)0}_{\;\;\;\;\;\;0} & = \sum_{m=0}^4 {\beta}_m^{(i,i+1)} \binom{3}{m}\left(\frac{a_{i+1}}{a_i}\right)^m + \sum_{m=0}^4 {\beta}_{D-m}^{(i-1,i)} \binom{3}{m}\left(\frac{a_{i-1}}{a_i}\right)^m \; ,
    \\
        W^{(i)j}_{\;\;\;\;\;\;k} &= \delta^j_k\Bigg\{ \bigg[{\beta}_0^{(i,i+1)} + {\beta}_1^{(i,i+1)}\left(\frac{c_{i+1}}{c_i} + 2\frac{a_{i+1}}{a_i}\right) 
        \notag\\
        &\qquad +{\beta}_2^{(i,i+1)}\left(2\frac{c_{i+1}}{c_i}\frac{a_{i+1}}{a_i} + \frac{a_{i+1}^2}{a_i^2}\right)
        + {\beta}_3^{(i,i+1)}\frac{c_{i+1}}{c_i}\frac{a_{i+1}^2}{a_i^2}\bigg]
        \notag\\
        &\qquad +\bigg[{\beta}_4^{(i-1,i)} + {\beta}_3^{(i-1,i)}\left(\frac{c_{i-1}}{c_i} + 2\frac{a_{i-1}}{a_i}\right) 
        \notag\\
        &\qquad +{\beta}_2^{(i-1,i)}\left(2\frac{c_{i-1}}{c_i}\frac{a_{i-1}}{a_i} + \frac{a_{i-1}^2}{a_i^2}\right)  + {\beta}_1^{(i-1,i)}\frac{c_{i-1}}{c_i}\frac{a_{i-1}^2}{a_i^2}\bigg]\Bigg\}  \; ,
\end{align}
where it is understood that $\beta_m^{(i,j)}$ are functions of the $b_i(t)$, as in Eq. \eqref{NewBetas}.

Lastly, the terms involving derivatives of $b_i$ are:
\begin{align}
    \frac{1}{b_i}\left[\nabla^{(i)0}\nabla^{(i)}_0 b_i - \delta^0_0\Box^{(i)}b_i\right] &= \frac{3}{c_i^2}\frac{\dot{a}_i}{a_i}\frac{\dot{b}_i}{b_i} \; ,
    \\
    \frac{1}{b_i}\left[\nabla^{(i)j}\nabla^{(i)}_k b_i - \delta^j_k\Box^{(i)}b_i\right] &= \frac{1}{c_i^2}\left(2\frac{\dot{a}_i}{a_i}\frac{\dot{b}_i}{b_i} + \frac{\ddot{b}_i}{b_i} - \frac{\dot{b}_i}{b_i}\frac{\dot{c}_i}{c_i}\right) \delta^j_k \; ,
\end{align}
and we note that the energy-momentum tensors are $T^{(i)}_{\mu\nu}=0$ for all sites in the bulk.

The continuum limit requires that we take $N\rightarrow\infty$ and $\dy\rightarrow0$ while keeping the product $N\dy=L$ fixed; this implies that we can Taylor expand all quantities living on sites $i\pm1$ around site $i$ with expansion parameter $\dy$ as:
\begin{equation}
    X_{i\pm1} = X_i \pm\dy X_i' + \frac12\dy^2 X_i'' \pm \mathcal{O}(\dy^3)\; .
\end{equation}

Let us look first at the 00-equation. Putting together the above expressions, substituting $M_{(4)}^2=M_{(5)}^3\dy$ and $\Lambda_4=\Lambda_5\dy$, as well as the explicit GR form of the $\beta_m^{(i,j)}$ coefficients from Eq. \eqref{NewBetas}, this equation reads:
\begin{equation}
\begin{split}
    -\frac{3M_{(5)}^3\dy}{c_i^2}&\left(\frac{\dot{a}_i}{a_i}\right)^2 + 2\Lambda_5\dy + \left\{\frac{2}{b_i(b_i+b_{i+1})}\frac{M_{(5)}^3}{\dy}\left[-6+9\frac{a_{i+1}}{a_i}-3\left(\frac{a_{i+1}}{a_i}\right)^2\right]\right\}
    \\
    &+ \left\{\frac{2}{b_i(b_i+b_{i-1})}\frac{M_{(5)}^3}{\dy}\left[-3\left(\frac{a_{i-1}}{a_i}\right)^2+3\left(\frac{a_{i-1}}{a_i}\right)^3\right]\right\} = \frac{3M_{(5)}^3\dy}{c_i^2}\frac{\dot{a}_i}{a_i}\frac{\dot{b}_i}{b_i} \; .
\end{split}
\end{equation}
Substituting in the Taylor expansions for $a_{i\pm1}$ and $b_{i\pm1}$, the first term in curly brackets becomes, to first order in $\dy$ (suppressing the $i$ indices for brevity):
\begin{equation}
    \{\text{Curly bracket 1}\} = \frac{M_{(5)}^3}{b^2}\left[3\frac{a'}{a} + \dy\left(-\frac32\frac{a'}{a}\frac{b'}{b} + \frac32\frac{a''}{a}-3\frac{a^{\prime2}}{a^2}\right)\right] \; ,
\end{equation}
and the second term becomes:
\begin{equation}
    \{\text{Curly bracket 2}\} = \frac{M_{(5)}^3}{b^2}\left[-3\frac{a'}{a} + \dy\left(-\frac32\frac{a'}{a}\frac{b'}{b} + \frac32\frac{a''}{a}+6\frac{a^{\prime2}}{a^2}\right)\right] \; .
\end{equation}
Adding the two together and taking the limit $\dy\rightarrow 0$, we get the continuum equation:
\begin{equation}
    \frac{3}{c^2}\left(\frac{\dot{a}^2}{a^2}+\frac{\dot{a}}{a}\frac{\dot{b}}{b}\right) + \frac{3}{b^2} \left(\frac{a'}{a}\frac{b'}{b} -\frac{a''}{a}-\frac{a^{\prime2}}{a^2}\right) = \frac{2\Lambda_5}{M_{(5)}^3} \; ,
\end{equation}
which is exactly Eq. \eqref{G00} for $G^0_{\;0}$.

Next up is the $jk$-equation, which in the deconstructed theory is as follows:
\begin{align}
    &\frac{M_{(5)}^3\dy}{c_i^2}\left(-\frac{\dot{a}^2_i}{a^2_i} - 2\frac{\ddot{a}_i}{a_i} + 2\frac{\dot{a}_i}{a_i}\frac{\dot{c}_i}{c_i}\right) +2\Lambda_5\dy - \frac{M_{(5)}^3\dy}{c_i^2}\left(2\frac{\dot{a}_i}{a_i}\frac{\dot{b}_i}{b_i} + \frac{\ddot{b}_i}{b_i} - \frac{\dot{b}_i}{b_i}\frac{\dot{c}_i}{c_i}\right)
    \\
\begin{split}
    &= \Bigg\{ \frac{2}{b_i(b_i+b_{i+1})}\frac{M_{(5)}^3}{\dy}\bigg[-6+3\left(\frac{c_{i+1}}{c_i}+2\frac{a_{i+1}}{a_i}\right)-\left(2\frac{c_{i+1}}{c_i}\frac{a_{i+1}}{a_i} + \frac{a_{i+1}^2}{a_i^2}\right)\bigg]\Bigg\}
    \notag\\
    &\qquad +\Bigg\{\frac{2}{b_i(b_i+b_{i-1})}\frac{M_{(5)}^3}{\dy}\bigg[3\left(\frac{c_{i-1}}{c_i}\frac{a_{i-1}^2}{a_i^2}\right) - \left(2\frac{c_{i-1}}{c_i}\frac{a_{i-1}}{a_i}+\frac{a_{i-1}^2}{a_i^2}\right)\bigg]\Bigg\} \; .
\end{split}
\end{align}
Performing our Taylor expansion on each of $a$, $b$ and $c$ this time, the first term in the curly brackets becomes:
\begin{equation}
    \{\text{Curly bracket 1}\} = \frac{M_{(5)}^3}{b^2}\left[\frac{c'}{c}+2\frac{a'}{a} + \dy\left(-\frac12\frac{b'}{b}\frac{c'}{c} -\frac{a'}{a}\frac{b'}{b} + \frac12\frac{c''}{c}-2\frac{a'}{a}\frac{c'}{c}-\frac{a^{\prime2}}{a^2}+\frac{a''}{a}\right)\right] \; ,
\end{equation}
while the second term becomes:
\begin{equation}
    \{\text{Curly bracket 2}\} = \frac{M_{(5)}^3}{b^2}\left[-\frac{c'}{c}-2\frac{a'}{a} + \dy\left(-\frac12\frac{b'}{b}\frac{c'}{c} -\frac{a'}{a}\frac{b'}{b} + \frac12\frac{c''}{c}+4\frac{a'}{a}\frac{c'}{c}+2\frac{a^{\prime2}}{a^2}+\frac{a''}{a}\right)\right] \; .
\end{equation}
Summing and taking $\dy\rightarrow0$, we arrive at the continuum equation:
\begin{equation}
\begin{split}
    &\frac{1}{c^2} \left(-\frac{\dot{a}^2}{a^2}+2\frac{\dot{a}}{a}\frac{\dot{c}}{c}-2\frac{\dot{a}}{a}\frac{\dot{b}}{b}+\frac{\dot{b}}{b}\frac{\dot{c}}{c}-2\frac{\ddot{a}}{a}-\frac{\ddot{b}}{b}\right) \\&+ \frac{1}{b^2} \left(\frac{a^{\prime2}}{a^2}-2\frac{a^\prime}{a}\frac{b^\prime}{b}+2\frac{a'}{a}\frac{c'}{c}-\frac{b'}{b}\frac{c'}{c}+2\frac{a''}{a}+\frac{c''}{c}\right) = -\frac{2\Lambda_5}{M_{(5)}^3} \; ,
\end{split}
\end{equation}
which again is exactly Eq. \eqref{Gij} for $G^i_{\;j}$. Thus, we recover \emph{all} of the dynamical field equations for brane cosmology in the extra dimensional bulk upon taking the continuum limit of the STMG field equations for the sites in the bulk of its theory graph.

\subsection{Israel junction conditions}

On the boundary sites, the corresponding metrics possess only a single interaction. At $y=0$, corresponding to the site $i=0$, this interaction is positively oriented, so leads in the continuum limit to the term we called `curly bracket 1' in the previous calculation; at $y=L$, corresponding to the site $i=N-1$, the converse is true, and the one negatively oriented interaction gives rise to `curly bracket 2'. 

With only one of these brackets present, the $\mathcal{O}(1)$ terms no longer cancel one another; they are instead compensated for by the energy-momentum tensors we have coupled to the boundary sites, of the form \eqref{perfect_fluid_Tmunu}. Therefore, in the limit $\dy\rightarrow0$, the 00-component of the STMG field equations \eqref{NewFieldEqs} on site $i=0$ becomes:
\begin{equation}
    \frac{3M_{(5)}^3}{b^2}\left.\frac{a'}{a}\right|_{i=0} = \frac{T^{(0)0}_{\;\;\;\;\;\;0}}{b} \implies 3M_{(5)}^3\left.\frac{a'}{ab}\right|_{y=0} = -(\rho_0 + \sigma_0) \; ,
\end{equation}
which is precisely the junction condition \eqref{K00} for $K^0_{\;0}$ at $y=0$. On the opposite boundary, we use get the desired orientation flip:
\begin{equation}
    -\frac{3M_{(5)}^3}{b^2}\left.\frac{a'}{a}\right|_{i=N-1} = \frac{T^{(N-1)0}_{\;\;\;\;\;\;\;\;\;\;\;\;0}}{b} \implies 3M_{(5)}^3\left.\frac{a'}{ab}\right|_{y=L} = +(\rho_L + \sigma_L) \; .
\end{equation}

Similarly, in the limit $\dy\rightarrow0$, the $jk$-components of the field equations on the site $i=0$ become:
\begin{equation}
    \frac{M_{(5)}^3}{b^2}\delta^j_k\left.\left(\frac{c'}{c}+2\frac{a'}{a}\right)\right|_{i=0} = \frac{T^{(0)j}_{\;\;\;\;\;\;k}}{b} \implies M_{(5)}^3\left.\left(\frac{c'}{cb}+2\frac{a'}{ab}\right)\right|_{y=0} = (p_0 - \sigma_0) \; ,
\end{equation}
which is precisely the junction condition \eqref{Kij} for $K^j_{\;k}$ at $y=0$. Again, one recovers the orientation flip on the opposite boundary:
\begin{equation}
    \frac{M_{(5)}^3}{b^2}\delta^j_k\left.\left(-\frac{c'}{c}-2\frac{a'}{a}\right)\right|_{i=N-1} = \frac{T^{(N-1)j}_{\;\;\;\;\;\;\;\;\;\;\;\;k}}{b} \implies M_{(5)}^3\left.\left(\frac{c'}{cb}+2\frac{a'}{ab}\right)\right|_{y=L} = -(p_L - \sigma_L) \; .
\end{equation}
Thus all of the Israel junction conditions for the branes in the continuum theory are encoded within the deconstructed theory via the field equations of the sites located at the endpoints of the chain of interactions.

\subsection{Momentum constraint}

Next up is the 5-dimensional momentum constraint, which should be encoded within the 4-dimensional Bianchi constraint, since it is equivalent to the equations of motion for the Stückelberg fields -- the analogues of the shift vector in the deconstructed theory. We recall that the STMG Bianchi constraint is (c.f. Eq. \eqref{NewBianchiConstraint}):
\begin{equation}
    \nabla^{(i)}_{\mu}\Wi{i} = \frac{M_{(4)}^2}{2}R_{(i)}\partial_\nu b_i \; .
\end{equation}

Since $b_i=b_i(t)$ depends only on time, the spatial components of the divergence of the $W$-tensors should vanish (since $\partial_j b_i=0$); with our cosmological ansatz \eqref{FRW} for the metrics, one can indeed show that $\nabla^{(i)}_\mu W^{(i)\mu}_{\;\;\;\;\;\;j}=0$ \emph{identically}. Therefore, all of the interesting behaviour resides in the time-component, which takes the form:
\begin{equation}
\begin{split}
    \nabla^{(i)}_\mu  W^{(i)\mu}_{\;\;\;\;\;\;0} &= 3\left(\frac{\dot{a}_{i+1}}{a_i} - \frac{c_{i+1}}{c_i}\frac{\dot{a}_i}{a_i}\right)\sigma_{i,i+1}^{(+)}- \frac{2(\dot{b}_i+\dot{b}_{i+1})}{(b_i+b_{i+1})^2}\sum_{m=0}^4 \bar{\beta}_m \binom{3}{m}\left(\frac{a_{i+1}}{a_i}\right)^m
    \\
    & + 3\left(\frac{\dot{a}_{i-1}}{a_i} - \frac{c_{i-1}}{c_i}\frac{\dot{a}_i}{a_i}\right)\sigma_{i,i-1}^{(-)} - \frac{2(\dot{b}_i+\dot{b}_{i-1})}{(b_i+b_{i-1})^2}\sum_{m=0}^4 \bar{\beta}_{4-m} \binom{3}{m}\left(\frac{a_{i-1}}{a_i}\right)^m \; ,
\end{split}
\end{equation}
where the $\sigma_{i,j}^{(\pm)}$ are defined, as in \cite{BH2}, to be:
\begin{align}
    \sigma^{(+)}_{i,j} &=  \sum_{m=0}^4 \beta_m^{(i,j)}\binom{2}{m-1}\left(\frac{a_{j}}{a_i}\right)^{m-1} \; ,\label{sig_p}
    \\
    \sigma^{(-)}_{i,k} &= \sum_{m=0}^4 \beta_{4-m}^{(k,i)}\binom{2}{m-1}\left(\frac{a_{k}}{a_i}\right)^{m-1} \; ,\label{sig_m}
\end{align}
and are related by:
\begin{equation}\label{Sig_pm}
     \sigma^{(-)}_{j,i} = \left(\frac{a_{i}}{a_{j}}\right)^{2} \sigma^{(+)}_{i,j} \; .
\end{equation}
We now proceed as we have been doing, by Taylor expanding the $a_{i\pm1}$, $b_{i\pm1}$ and $c_{i\pm1}$ around site $i$. To lowest order, one arrives at the following equation:
\begin{equation}
    \frac{3M_{(5)}^3}{b^2}\left(\frac{\dot{a}'}{a} - \frac{\dot{a}}{c}\frac{c'}{c}-\frac{\dot{b}}{b}\frac{a'}{a}\right) - \frac{3M_{(5)}^3}{b^2}\left(\frac{\dot{a}'}{a} - \frac{\dot{a}}{c}\frac{c'}{c}-\frac{\dot{b}}{b}\frac{a'}{a}\right) + \mathcal{O}(\dy) = 0 \; ,
\end{equation}
where the first bracket comes from the positive orientation interaction and the second bracket comes from the negative orientation interaction. Obviously, in the bulk -- where both of these interactions are present simultaneously -- the above equation is satisfied trivially. However, as we just saw, on the boundary sites only one of these two terms is present, so in order to satisfy the Bianchi constraint for every site, the term in brackets must \emph{itself} vanish. This implies that:
\begin{equation}
    \frac{3M_{(5)}^3}{b^2}\left(\frac{\dot{a}'}{a} - \frac{\dot{a}}{c}\frac{c'}{c}-\frac{\dot{b}}{b}\frac{a'}{a}\right) = 0 \; ,
\end{equation}
which is of course the 5-dimensional momentum constraint $G^5_{\;0}=0$, given in Eq. \eqref{G05}. So, we have shown that the Bianchi constraint in 4-dimensions encodes the continuum equation of the shift vector in 5-dimensions.

\subsection{Hamiltonian constraint}

Finally, we need to show that the scalar equations (c.f. Eq. \eqref{ScalarConstraint}):
\begin{equation}
    \frac{M_{(4)}^2}{2} R_{(i)} - 2\Lambda_4 = \sum_{m=0}^4 \frac{\partial\beta_m^{(i,i+1)}}{\partial b_i}e_m(S_{i\rightarrow i+1}) + \sum_{m=0}^4 \frac{\partial\beta_{4-m}^{(i-1,i)}}{\partial b_i}e_m(S_{i\rightarrow i-1}) \; ,
\end{equation}
become the 5-dimensional Hamiltonian constraint upon taking the continuum limit. This is the piece of information that was missing in standard multi-gravity but is now present in STMG.

With our cosmological metric ansatz \eqref{FRW}, the quantities we need are:
\begin{align}
    R_{(i)} &= \frac{6}{c_i^2}\left(\frac{\ddot{a}_i}{a_i}+\frac{\dot{a}_i^2}{a_i^2} - \frac{\dot{a}_i}{a_i}\frac{\dot{c}_i}{c_i}\right) \; ,
    \\
    e_0(S_{i\rightarrow j}) &= 1 \; ,
    \\
    e_1(S_{i\rightarrow j}) &= \frac{c_j}{c_i} + 3\frac{a_j}{a_i} \; ,
    \\
    e_2(S_{i\rightarrow j}) &= 3\left(\frac{c_j}{c_i}\frac{a_j}{a_i} + \frac{a_j^2}{a_i^2}\right) \; ,
    \\
    e_3(S_{i\rightarrow j}) &= 3\frac{c_j}{c_i}\frac{a_j^2}{a_i^2} + \frac{a_j^3}{a_i^3} \; ,
    \\
    e_4(S_{i\rightarrow j}) &= \frac{c_j}{c_i}\frac{a_j^3}{a_i^3} \; .
\end{align}
Substituting these into our scalar constraints, along with the GR form of the $\beta_m^{(i,j)}$ coefficients given in Eq. \eqref{NewBetas}, the explicit expression is:
\begin{equation}
    \begin{split}
        &0 = \frac{3M_{(5)}^3\dy}{c_i^2} \left(\frac{\ddot{a}_i}{a_i}+\frac{\dot{a}_i^2}{a_i^2} - \frac{\dot{a}_i}{a_i}\frac{\dot{c}_i}{c_i}\right) - 2\Lambda_5\dy
        \\
        &+ \Bigg\{\frac{2}{(b_i+b_{i+1})^2}\frac{M_{(5)}^3}{\dy}\bigg[-6+3\left(\frac{c_{i+1}}{c_i}+3\frac{a_{i+1}}{a_i}\right) -3\left(\frac{c_{i+1}}{c_i}\frac{a_{i+1}}{a_i} + \frac{a_{i+1}^2}{a_i^2}\right)\bigg]\Bigg\}
        \\
        &+ \Bigg\{\frac{2}{(b_i+b_{i-1})^2}\frac{M_{(5)}^3}{\dy}\bigg[-3\left(\frac{c_{i-1}}{c_i}\frac{a_{i-1}}{a_i} + \frac{a_{i-1}^2}{a_i^2}\right) +3\left(3\frac{c_{i-1}}{c_i}\frac{a_{i-1}^2}{a_i^2} + \frac{a_{i-1}^3}{a_i^3}\right) -6 \frac{c_{i-1}}{c_i}\frac{a_{i-1}^3}{a_i^3}\bigg]\Bigg\} \; .
    \end{split}
\end{equation}
Taylor expanding, one finds that both curly brackets contribute the same amount, which is, to first order in $\dy$:
\begin{equation}
    \{\text{Both curly brackets}\} = \dy\left(-\frac32\frac{a'}{a}\frac{c'}{c}-\frac32\frac{a^{\prime2}}{a^2}\right) \; ,
\end{equation}
so upon adding them up and taking $\dy\rightarrow0$, one arrives at last at the continuum equation:
\begin{equation}
    \frac{3}{b^2}\left( \frac{a^{\prime2}}{a^2}+\frac{a'}{a}\frac{c'}{c}\right) + \frac{3}{c^2}\left(\frac{\dot{a}}{a}\frac{\dot{c}}{c}-\frac{\dot{a}^2}{a^2}-\frac{\ddot{a}}{a}\right) = -\frac{2\Lambda_5}{M_{(5)}^3} \; ,
\end{equation}
which is precisely the 5-dimensional Hamiltonian constraint, Eq. \eqref{G55}. Everything checks out exactly as it should: we recover brane cosmology in its entirety in the continuum limit of STMG (with the interaction coefficients tuned to lead to pure GR in 5-dimensions). It seems that this STMG model really is the theory that arises in 4-dimensions upon deconstructing 5-dimensional GR!

\section{Einstein frame form of scalar-tensor multi-gravity}\label{app:Einstein frame}

In complete generality, the Jordan frame action of any STMG theory may be written as:
\begin{align}
    I &= I_K + I_V + I_M\left[\psi_i; g_{(i)}\right]
    \\
    I_K &= \sum_{i=0}^{N-1} \int \dd[D]x\sqrt{-\det g_{(i)}} \, \left[\frac{F_i(\phi_i)}{2}R_{(i)} - \frac{G_{i}(\phi_i)}{2}g^{(i)\mu\nu}\partial_\mu\phi_i\partial_\nu\phi_i - U_i(\phi_i)\right]
    \\
    I_V &= -\sum_{i,j} \int \dd[D]x \sqrt{-\det g_{(i)}} \sum_{m=0}^{D} \beta_m^{(i,j)}\left(\bm{\phi}\right) e_m(S_{i\rightarrow j})  \; ,
\end{align}
The notation for the matter action $I_M\left[\psi_i; g_{(i)}\right]$ means that the matter fields, $\psi_i$, minimally couple to their corresponding metrics $\gi{i}$, in this frame.

One may move into the Einstein frame by making the following conformal transformations on the various metrics \cite{Sergio}:
\begin{equation}
    \gi{i} \rightarrow \left(\frac{\tilde{M}_i^2}{F_i(\phi_i)}\right)^{\frac{D-2}{2}}\tilde{g}^{(i)}_{\mu\nu} \equiv  A_i^2(\phi_i)\tilde{g}^{(i)}_{\mu\nu} \; ,
\end{equation}
for which the action becomes:
\begin{align}
    I &= I_K + I_V + I_M\left[\psi_i; A_i^2(\phi_i)\tilde{g}_{(i)}\right]\label{Einstein frame action}
    \\
    I_K &= \sum_{i=0}^{N-1} \int \dd[D]x\sqrt{-\det \tilde{g}_{(i)}} \, \left[\frac{\tilde{M}_i^{D-2}}{2}\tilde{R}_{(i)} - \frac{\tilde{G}_{i}(\phi_i)}{2}\tilde{g}^{(i)\mu\nu}\partial_\mu \phi_i \partial_\nu\phi_i - \tilde{V}_i(\phi_i)\right]
    \\
    I_V &= -\sum_{i,j} \int \dd[D]x \sqrt{-\det \tilde{g}_{(i)}} \sum_{m=0}^{D} \tilde{\beta}_m^{(i,j)}\left(\bm{\phi}\right) e_m(\tilde{S}_{i\rightarrow j})  \; ,
\end{align}
where one has:
\begin{equation}
    \tilde{G}_i(\phi_i) = \tilde{M}_i^{D-2} \left[\frac{G_i(\phi_i)}{F_i(\phi_i)}+\frac{D-1}{D-2}\frac{F_i'(\phi_i)^2}{F_i(\phi_i)^2}\right] \; ,
\end{equation}
and:
\begin{equation}
    \tilde{V}_i(\phi_i) = A_i^{D}(\phi_i) U_i(\phi_i) \; .
\end{equation}
Lastly, under these conformal transformations, the $S_{i\rightarrow j}$ matrices become:
\begin{equation}
    S_{i\rightarrow j} = \frac{A_j(\phi_j)}{A_i(\phi_i)} \tilde{S}_{i\rightarrow j} \; ,
\end{equation}
which leads to the rescaling $\beta_m^{(i,j)}\rightarrow\tilde{\beta}_m^{(i,j)}$ given by:
\begin{equation}
    \tilde{\beta}_m^{(i,j)}(\bm{\phi}) = A_i^{D-m}(\phi_i)A_j^{m} (\phi_j)\beta_m^{(i,j)}(\bm{\phi}) \; .
\end{equation}
We stress that all the tilded quantities written throughout this appendix bear no relation to those we introduced when applying the Stückelberg trick in section \ref{Sec:stuckelberg}; here, they simply distinguish quantities in the Einstein frame from quantities in the Jordan frame.

In order to work with the Einstein frame action \eqref{Einstein frame action}, it helps to canonically normalise the scalar fields, which we can do by solving the integrals:
\begin{equation}
    \tilde{\chi}_i = \int_0^{\phi_i} \dd \hat{\phi}_i \,\sqrt{\tilde{G}_i(\hat{\phi}_i)} \; ,
\end{equation}
finally taking our action into its canonical form:
\begin{align}
    I &= I_K + I_V + I_M\left[\psi_i; A_i^2(\tilde{\chi}_i)\tilde{g}_{(i)}\right]\label{Einstein frame action canonical norm}
    \\
    I_K &= \sum_{i=0}^{N-1} \int \dd[D]x\sqrt{-\det \tilde{g}_{(i)}} \, \left[\frac{\tilde{M}_i^{D-2}}{2}\tilde{R}_{(i)} - \frac{1}{2}\tilde{g}^{(i)\mu\nu}\partial_\mu \tilde{\chi}_i \partial_\nu\tilde{\chi}_i - \tilde{V}_i(\tilde{\chi}_i)\right]
    \\
    I_V &= -\sum_{i,j} \int \dd[D]x \sqrt{-\det \tilde{g}_{(i)}} \sum_{m=0}^{D} \tilde{\beta}_m^{(i,j)}\left(\tilde{\bm{\chi}}\right) e_m(\tilde{S}_{i\rightarrow j})  \; .
\end{align}
The field equations arising from the Einstein frame action \eqref{Einstein frame action canonical norm} are:
\begin{equation}\label{Einstein frame eqs}
\boxed{
    \tilde{M}_i^{D-2}\tilde{G}^{(i)}_{\mu\nu} + \tilde{W}^{(i)}_{\mu\nu}(\tilde{\bm{\chi}}) = A_i^2(\tilde{\chi}_i)T^{(i,m)}_{\mu\nu} + \tilde{T}^{(i,\tilde{\chi}_i)}_{\mu\nu} } \; ,
\end{equation}
by varying with respect to the metrics, where the $\tilde{W}$-tensor is given by Eq. \eqref{STMG W} after replacing $\beta_m^{(i,j)}(\bm{\phi})$ by $\tilde{\beta}_m^{(i,j)}(\tilde{\bm{\chi}})$. The matter energy-momentum tensor has no tilde, as it is actually defined with respect to the \emph{Jordan frame} metric by:
\begin{equation}
    T^{(i,m)}_{\mu\nu} = \frac{-2}{\sqrt{-\det g_{(i)}}}\frac{\delta I_M}{\delta g^{(i)\mu\nu}} = -2\frac{\partial\mathcal{L}_m}{\partial g^{(i)\mu\nu}} + g^{(i)}_{\mu\nu}\mathcal{L}_m \; ;
\end{equation}
this is because matter still couples minimally to $\gi{i}=A_i^2(\tilde{\chi}_i)\tilde{g}^{(i)}_{\mu\nu}$ rather than to $\tilde{g}^{(i)}_{\mu\nu}$ (see e.g. \cite{Brax_lectures} for more details). The scalars, on the other hand, have their kinetic terms coupled minimally to the Einstein frame metrics, so contribute the following energy-momentum tensors:
\begin{equation}
    \tilde{T}^{(i,\tilde{\chi}_i)}_{\mu\nu} = \partial_\mu\tilde{\chi}_i\partial_\nu\tilde{\chi}_i - \frac12\tilde{g}^{(i)}_{\mu\nu}\partial_\lambda\tilde{\chi}_i \partial^\lambda\tilde{\chi}_i - \tilde{g}^{(i)}_{\mu\nu} \tilde{V}_i(\tilde{\chi}_i) \; .
\end{equation}
The field equations \eqref{Einstein frame eqs} are to be compared against the corresponding Jordan frame expression, Eq. \eqref{STMG field eqs}.

In the scalar sector, the field equations are:
\begin{equation}\label{Einstein frame scalar eq}
\boxed{
    \tilde{\Box}^{(i)}\tilde{\chi}_i = \frac{\partial\tilde{V}_i}{\partial\tilde{\chi}_i} + \tilde{\mathbb{X}}_i(\tilde{\bm{\chi}}) - A_i^3(\tilde{\chi}_i)\frac{\partial A_i}{\partial\tilde{\chi}_i} T_{(i,m)} } \; ,
\end{equation}
where $T_{(i,m)}$ are the traces of the Jordan frame matter energy-momentum tensors. The term $\tilde{\mathbb{X}}_i(\tilde{\bm{\chi}})$ is given by Eq. \eqref{metric X} after making the replacements $\beta_m^{(i,j)}(\bm{\phi})\rightarrow\tilde{\beta}_m^{(i,j)}(\tilde{\bm{\chi}})$, $S_{i\rightarrow j}\rightarrow \tilde{S}_{i\rightarrow j}$ and $\phi_i\rightarrow\tilde{\chi}_i$. This Einstein frame scalar equation is to be compared with Eq. \eqref{STMG scalar dynamics} in the Jordan frame, which is far more complicated.

If one wishes to model build within the framework of STMG, the canonically normalised Einstein frame action \eqref{Einstein frame action canonical norm} and its corresponding field equations \eqref{Einstein frame eqs} and \eqref{Einstein frame scalar eq}, with some choices for the scalar potentials $\tilde{V}_i(\tilde{\chi}_i)$ and interaction coefficients $\tilde{\beta}_m^{(i,j)}(\tilde{\bm{\chi}})$, probably constitute the simplest place to start. However, by using the more complicated Jordan frame form of the action that we used throughout the main paper, given in Eq. \eqref{STMG action}, the link to higher dimensional gravity and dimensional deconstruction (in the situations where a continuum limit may be defined) is manifest, as the Jordan frame scalars correspond to the value of the extra dimensional lapse function on different hypersurfaces -- see section \ref{Sec:new deconstruction} for a recap of how this works. This intuition about the scalars is sacrificed in the Einstein frame in favour of calculational ease.

To close, we note that in the bi-metric Starobinsky model first introduced via its Einstein frame action in refs. \cite{Starobinsky_bigravity,Ghost_freedom_quadratic_bigravity}, which we cited in sections \ref{Sec:new deconstruction} and \ref{Sec:MMST} as an example of an STMG model that has previously been studied in an inflationary context, the action, Einstein equations and scalar field equations indeed take the form of Eqs. \eqref{Einstein frame action canonical norm}, \eqref{Einstein frame eqs} and \eqref{Einstein frame scalar eq}, as they should.

\bibliography{bibliography.bib}
\bibliographystyle{JHEP}
\end{document}